\documentclass[fleqn,usenatbib]{mnras}

\usepackage[T1]{fontenc}

\DeclareRobustCommand{\VAN}[3]{#2}
\let\VANthebibliography\thebibliography
\def\thebibliography{\DeclareRobustCommand{\VAN}[3]{##3}\VANthebibliography}

\usepackage{graphicx}	
\usepackage{amsmath}	
\usepackage{amssymb}	
\usepackage{pdflscape}
\usepackage[bottom]{footmisc}

\title[Probing TTVs of Hot-Jupiters ] {Probing Transit Timing Variations of Three Hot-Jupiters: HATP-36b, HATP-56b, and WASP-52b}

\author[Sonbas et al.]{E. Sonbas,$^{1,2}$\thanks{E-mail: edasonbas@gmail.com}
N. Karaman,$^{3}$
A. \"Ozd\"onmez,$^{4}$
H. Er,$^{4}$
K. S. Dhuga,$^{2}$
E. G\"o\u{g}\"u\c{s},$^{5}$, 
I. Nasiroglu $^{4}$ 
\newauthor and M. Zejmo $^{6}$
\\
$^{1}$Department of Physics, Adiyaman University, 02040 Adiyaman, Turkey\\
$^{2}$ Department of Physics, The George Washington University, Washington, DC 20052, USA\\
$^{3}$Department of Electric Electronic Engineering, Adiyaman University, 02040 Adiyaman, Turkey \\
$^{4}$Departments of Astronomy and Astrophysics, Ataturk University Yakutiye, 25240, Erzurum, Turkey\\
$^{5}$Sabanc\i~University, Orhanl\i~- Tuzla, Istanbul 34956, Turkey \\
$^{6}$Janusz Gil Institute of Astronomy, University of Zielona Gora, Prof. Szafrana 2, PL-65-516 Zielona Gora, Poland
}

\date{Accepted 2021 November 6. Received 2021 October 21; in original form 2021 March 10}

\pubyear{20XX}

\begin{document}
\label{firstpage}
\pagerange{\pageref{firstpage}--\pageref{lastpage}}
\maketitle

\begin{abstract}
We report the results of new transit observations for the three hot Jupiter-like planets HATP-36b, HATP-56b and WASP-52b respectively. Transit timing variations (TTVs) are presented for these systems based on observations that span the period 2016 - 2020. The data were collected with the 0.6 m telescope at Adiyaman University (ADYU60, Turkey) and the 1.0 m telescope at TÜBİTAK National Observatory (TUG, Turkey). Global fits were performed to the combined light curves for each system along with the corresponding radial velocity (RV) data taken from the literature. The extracted parameters (for all three systems) are found to be consistent with the values from previous studies. Through fits to the combined mid-transit times data from our observations and the data available in the literature, an updated linear ephemeris is obtained for each system. Although a number of potential outliers are noted in the respective O-C diagrams, the majority of the data are consistent within the 3$\sigma$ confidence level implying a lack of convincing evidence for the existence of additional objects in the systems studied.
\end{abstract}




\section{Introduction}
Since the discovery of the first hot Jupiter-like exoplanet \citep{1995Natur.378..355M} several thousand exoplanets ($\sim$ 4375) have been confirmed, the majority of which ($\sim$ 3354\footnote{$https://exoplanetarchive.ipac.caltech.edu/docs/counts$\_$detail.html$}) have been detected using the transit method. The light curves measured by the transit method not only verify or improve the parameters of the system, but also provide important information about the presence of additional objects in the system via the long-term transit time variations (TTVs; \cite{2005MNRAS.359..567A}).\\ 
\\
Discovered by the HATNet Exoplanet survey  \citep{2004PASP..116..266B}, HATP-36b, is a transiting hot-Jupiter (1.832M$_J$, 1.264R$_J$) orbiting around a V = 12.5 mag, G5V type Sun-like star (1.022M$_\odot$, 1.096R$_\odot$). This system has a short orbital period of 1.33 days \citep{2012AJ....144...19B, 2015AandA...579A.136M}. Following the discovery of HATP-36b, \citet{2013IBVS.6082....1M} followed the object with photometric observations using a 0.6 m telescope and improved the transit ephemeris.  The system has also been studied by \cite{2015AandA...579A.136M} as part of the GAPS program; the addition of new spectroscopic and photometric data has produced a revised set of physical parameters. The authors also noted anomalies in three photometric transit light curves and confirmed, by the analysis of the HARPS-N spectra, that these were compatible with star spot complexes on the surface of the star.  \cite{2019AJ....157...82W} combined 26 new transit observations for HATP-36b with the ones from the literature and investigated the existence of a third object. Despite the new data set, which covers a relatively long time period, they did not detect significant variations in the  transit timing variation (TTV) signal. \cite{2019AJ....158...39C} analyzed high-quality photometric data for HATP-36b using a wavelet-based denoising technique to minimize the effects of various background contributions such varying sky transparency and stellar activity. They presented updated results for the transit parameters with higher precision compared to the previous ones.\\
\\
HATP-56b, also discovered by the HATNet Exoplanet survey \citep{2015AJ....150...85H}, is an inflated hot-Jupiter exoplanet with the following physical parameters determined by combining ground-based photometric and spectroscopic observations: mass of M$_P$= 2.18M$_J$, and a radius of R$_P$=1.43R$_J$. It orbits a F-type star with magnitude of V = 10.91 and has an orbital period of 2.7908 days. The maximum rotational period for HatP-56 has been reported as 1.8$\pm$0.2 days based on the spectroscopic measurements  \citep{2016MNRAS.460.3376Z}. In addition, the periodogram of the K2 light curve suggests that the host star in the system is a $\gamma$ Dor variable \citep{2015AJ....150...85H}. Using Doppler tomographic analyses for the spectroscopic transits, \citet{2016MNRAS.460.3376Z} reported a spin–orbit aligned (8$^o$ $\pm$ 2$^o$) geometry for HATP-56b.  They also examined a tidal re-alignment model for rapidly rotating host stars similar to HATP-56b and found no evidence that the rotation rates of the system were altered by star–planet tidal interactions. Furthermore, they noted that HATP-56b is a system in which the rotation period of the host star is faster than the orbital period of the planet.\\
\\
WASP-52b is another planet that falls in the hot-Jupiter category and orbits a Sun-like K2V type star with a mass of 0.87($\pm$ 0.03)M$_\odot$ and radius of 0.79($\pm$ 0.02)R$_\odot$. It was first reported by \cite{2013AandA...549A.134H}. They estimated the physical parameters of the system as; planet mass  M$_P$= 0.46$\pm$0.02M$_J$, a planetary radius of R$_P$=1.27$\pm$0.03R$_J$, and an orbital period of  P = 1.7497798 $\pm$ 0.0000012 days. WASP-52 system has been studied by many authors to probe the anomalies seen in the transit light curves (e.g. potential effects due to stellar spots and/or bright regions in the stellar photosphere) \citep{2015JATIS...1b7002S, 2015MNRAS.450.3101B, 2016MNRAS.463.2922K, 2017AandA...600L..11C, 2017MNRAS.465..843M, 2017MNRAS.470..742L, 2018AJ....156..124B, 2019MNRAS.486.2290O}. A detailed TTV study by \citet{2015MNRAS.450.3101B} using the data collected from the literature did not show any indication of a third object in the system. Later studies suggested that the variations seen in the transit times could either be due to existence of a third object in the system or the effect of stellar spots \citep{2017MNRAS.465..843M, 2013MNRAS.430.3032B, 2013AandA...556A..19O, 2016AandA...585A..72I}. The TTVs have been investigated with the use of linear and quadratic models; parabolic changes in residuals of the transits have beeen reported by \citet{2019MNRAS.486.2290O}. In another study \citep{2017AandA...600L..11C}, spectroscopic observations suggest a possible cloudy atmosphere for WASP-52b.\\ 
\\
The possibility of detecting additional objects through the study of TTVs, and in particular, the intriguing notion of observing features in the measured light curves that might be attributable to stellar surface activity, provide the primary motivation for this work. In this study, we present the results of new transit observations for the three hot Jupiter-like planets HATP-36b, HATP-56b and WASP-52b respectively; the period of observations spans 2016 - 2020. The paper is organized as follows: in Section 2, we describe the photometric observations and reduction of the data acquired for HATP-36b, HATP-56b and WASP-52b and the details of the computation of the system parameters, and the TTV analysis. The implications of these results are discussed in Section 3. We conclude by summarizing our main findings in Section 4.  
\section{Observations and Data Reduction}
\subsection{Photometric Observations}\label{sec:phot_obs} 

New photometric observations of HATP-36b, HATP-56b and WASP-52b were carried out with 0.6 m and 1 m telescopes. The 0.6 m telescope (ADYU60) is located at Adiyaman University, Turkey and is currently operated remotely from the Adiyaman University Observatory. ADYU60 is equipped with 1k x 1k Andor iKon-M934 CCD with a pixel size and image scale of 13$\mu$m $\times$ 13$\mu$m and 0.67$\arcsec$/pixel respectively. The 1.0 m Ritchey–Chr\'etien (RC) telescope (T100) is located at the Bakirlitepe Mountain and is currently operated remotely from TUG (TÜBİTAK National Observatory) in Antalya, Turkey. The T100 telescope houses a 4k $\times$ 4k SI 1100 CCD camera operating at -90$^{\circ}$C. The pixel size, overall field of view and image scale of the CCD camera are 15$\mu$m $\times$ 15$\mu$m, $21.5\arcmin\times21.5\arcmin$ and 0.31$\arcsec$/pixel respectively. Photometric observations for Adyu60 were made with the $Johnson$ R filter and for T100 we used the $Bessel$ R filter. Defocussing technique was adopted in the T100 observations to improve the statistics of the photometric data for a given exposure time.\\
\\  
 We observed a total of sixteen transits for HATP-36b with Adyu60 and one transit with T100 between April 2016 - June 2020. In addition, Adyu60 was used to observe nine transits for HATP-56b between December 2016 - February 2018 and thirteen transits for WASP-52b between August 2016 - October 2020. In order to maintain relatively high signal-to-noise ratios, some exposure times were adjusted based on changing weather conditions and according to the magnitude of the target. However, during all phases of the transits, the exposure was kept fixed to acquire data with consistent time intervals. The summaries of our observations are given in Table \ref{tab:log_of_observations}.
\begin{table*}
\caption{Log of observations.}
\label{tab:log_of_observations}
\begin{tabular}{lccccccc}
\hline\hline 
Date of & Object  & Telescope & Filter & Exposure (s) & Number of & Airmass & RMS  \\
observations & & & & & data points& & mmag \\
\hline
01.04.2016 & HATP-36b & TUG & R & 45 & 164 & 1.09 - 1.80 &7.042\\
01.04.2016 & HATP-36b & ADYU60 & R & 120 & 79 & 1.13 – 1.76 & 4.667\\
11.05.2016 & HATP-36b & ADYU60 & R & 150 & 38 & 1.10  – 1.38 & 2.452\\
22.01.2017 & HATP-36b & ADYU60 & R & 60 & 133 & 1.24 – 1.06 & 4.728\\
19.02.2017 & HATP-36b & ADYU60 & R & 90 & 79 & 1.53 - 1.01 & 6.919\\
23.02.2017 & HATP-36b & ADYU60 & R & 90 & 87 & 1.57 – 1.01 & 6.050\\
25.04.2017 & HATP-36b & ADYU60 & R & 120 & 99 & 1.06 – 1.69 & 3.335\\
08.06.2017 & HATP-36b & ADYU60 & R & 120 & 73 & 1.01 – 1.21 & 5.758\\
30.01.2018 & HATP-36b & ADYU60 & R & 120 & 123 & 1.31 – 1.02 & 6.382\\
07.02.2018 & HATP-36b & ADYU60 & R & 120 & 165 & 1.39 – 1.04 & 13.329\\
30.03.2020 & HATP-36b & ADYU60 & R & 120 & 138 & 1.61 – 1.01 & 2.994\\
10.05.2020 & HATP-36b & ADYU60 & R & 120 & 119 & 1.08 – 2.18 & 6.172 \\
30.05.2020 & HATP-36b & ADYU60 & R & 120 & 138 & 1.03 – 2.00 & 3.541\\
03.06.2020 & HATP-36b & ADYU60 & R & 120 & 131 & 1.02 – 1.87 & 2.568\\
07.06.2020 & HATP-36b & ADYU60 & R & 120 & 111 & 1.03 – 1.67 & 5.978\\
11.06.2020 & HATP-36b & ADYU60 & R & 120 & 105 & 1.04 – 1.65 & 3.980\\
15.06.2020 & HATP-36b & ADYU60 & R & 120 & 91 & 1.05 - 1.57 & 2.989\\
\hline
08.12.2016 & HATP-56b & ADYU60 & R & 120 & 128 & 2.27 - 1.03 & 2.122\\
19.01.2017 & HATP-56b & ADYU60 & R & 120 & 112 & 2.02 - 1.05 & 2.353\\
13.02.2017 & HATP-56b & ADYU60 & R & 120 & 130 & 1.05 - 1.35 & 2.414\\
27.02.2017 & HATP-56b & ADYU60 & R & 120 & 107 & 1.07 - 1.17 & 2.217\\
22.10.2017 & HATP-56b & ADYU60 & R & 120 & 119 & 1.75 - 1.02 & 1.950\\
14.12.2017 & HATP-56b & ADYU60 & R & 120 & 143 & 1.02 - 1.90 & 2.748\\
28.12.2017 & HATP-56b & ADYU60 & R & 120 & 115 & 1.02 - 1.62 & 6.773\\
11.01.2018 & HATP-56b & ADYU60 & R & 120 & 74 & 1.02 - 1.56 & 3.825\\
08.02.2018 & HATP-56b & ADYU60 & R & 120 & 73 & 1.02 - 1.03 & 4.652\\
\hline
12.08.2016 & WASP-52b & ADYU60 & R & 120 & 82 & 1.26 - 1.18 & 2.931\\
26.08.2016 & WASP-52b & ADYU60 & R & 120 & 85 & 1.17 - 1.28 & 3.505\\
07.10.2016 & WASP-52b & ADYU60 & R & 120 & 95 & 1.23 - 2.8 & 4.241\\
22.01.2017 & WASP-52b & ADYU60 & R & 60 & 57 & 1.52 - 3.27 & 13.450\\
22.09.2017 & WASP-52b & ADYU60 & R & 120 & 55 & 1.15 - 1.40 & 6.421\\
13.10.2017 & WASP-52b & ADYU60 & R & 120 & 100 & 1.16 - 2.05 & 3.425\\
20.10.2017 & WASP-52b & ADYU60 & R & 120 & 90 & 1.20 - 2.30 & 4.483\\
27.10.2017 & WASP-52b & ADYU60 & R & 120 & 85 & 1.22 - 2.30 & 5.092\\
07.09.2018 & WASP-52b & ADYU60 & R & 120 & 90 & 1.41 - 1.17 &3.339\\
14.09.2018 & WASP-52b & ADYU60 & R & 120 & 90 & 1.33 - 1.19 & 3.164\\
05.10.2018 & WASP-52b & ADYU60 & R & 60 & 98 & 1.20 - 1.35 & 3.294\\
18.09.2020 & WASP-52b & ADYU60 & R & 120 & 94 & 2.17 - 1.16 & 4.601\\
02.10.2020 & WASP-52b & ADYU60 & R & 120 & 96 & 1.84 - 1.15 & 6.221\\
\hline
\end{tabular}
\end{table*}
\subsection{Data Reduction}\label{sec:data_red} 
 The software package, AstroImageJ (AIJ) (Collins et al. 2017), was deployed to perform data reduction, calibrations, extraction of differential aperture photometry and detrending parameters. AIJ is a powerful tool for image processing and precise photometry especially for exoplanet transit light curves. We used median-combined bias and flat frames to correct the raw CCD images by using the Data Processor module of AIJ. Dark correction was not applied because dark counts were negligible in the frames for both telescopes. We performed differential photometry using the Multi-Aperture (MA) module in AIJ. To obtain differential magnitudes for each system, we considered as many comparison stars as we could find in the CCD frames. In the final analysis, we selected two or three comparison stars with the least-variable light curves. The same set of standard stars were used for all observations of a given source. Photometric uncertainties, CCD read-out noise etc., were estimated using MA module of AIJ. The details of the selected comparisons are given in Table \ref{tab:comparisons}. In order to minimize the transit modeling residuals, we determined the aperture sizes of the target and comparison stars by using the "radial profile" feature of the AIJ program, and we allowed the aperture sizes to vary by 1.2 times the FWHM value in each image. The conversion from Julian Date (JD) to Barycentric Julian Date (BJD) was done within AIJ. We used AIJ Multi-plot module to extract detrend parameters. These parameters are airmass, time, sky background, FWHM of the average PSF in each image, total comparison star counts, and target x-centroid and y-centroid positions on the detector. These parameters, along with relative flux and corresponding flux errors, were then used to create simultaneous detrended light curves in EXOFASTv2 as part of the global and individual fits for consistency. In EXOFASTv2 we simply include additional columns of detrending parameters for each transit in the transit file to detrend against. We kept the same number of detrend parameters for each source. We used additive detrending scheme of EXOFASTv2 in our analysis. The final light curves were obtained from the differential magnitudes, and for each light curve, the RMS was calculated to obtain a measure of the quality of the data. The RMS varies in the range $\sim (1.9 - 7.0)$ \textit{mmags}, primarily reflecting the count rates and the different observing conditions during which the light curves were measured. 
\begin{table*}
\caption{The information on comparison stars. Magnitudes are taken from the SIMBAD catalog$^{1}$.}
\label{tab:comparisons}
\begin{tabular}{lllll}
\hline\hline 
Object & RA  & DEC & Kmag &  \\
\hline
HATP-36 & 12 33 03.909 & +44 54 55.180 & 10.603 & \\
2MASS 12331133+4451365 & 12 33 11.339 & +44 51 36.550 & 11.462 & \\
2MASS 12331452+4449494 & 12 33 14.528 & +44 49 49.490 & 12.172 & \\
2MASS 12330477+4457439 & 12 33 04.779 & +44 57 43.934 & 11.288 & \\
\hline

HATP-56 & 06 43 23.529 & +27 15 08.218 & 9.830 & \\
TYC 1901-762-1 & 06 43 21.915 & +27 16 33.245 & 11.734 & \\
TYC 1901-1083-1 & 06 43 21.789 & +27 17 50.300 & 11.505 & \\
\hline
WASP-52 & 23 13 58.76 & +08 45 40.6 & 10.086 & \\
2MASS 23140483+0844176 & 23 14 05.034 & +08 44 19.54 & 11.883 & \\
2MASS 23135436+0848293 & 23 13 54.44 & +08 48 31.48 & 12.087 & \\
\hline
$^{1} {http://simbad.u-strasbg.fr/simbad/}$
\end{tabular}
\end{table*}
\subsection{Global Fits}\label{sec:exofast} 
To determine the planetary system parameters, along with their uncertainties, we deployed the Markov Chain Monte Carlo code EXOFASTv2 \citep{2013PASP..125...83E, Eastman2017, Eastman2019}. The updated version of the code provides a general solution encompassing a multi-parameter space covering an arbitrary number of light curves along with multiple radial velocity (RV) and SED data sets. The code calculates both the stellar and planetary parameters as well as the transit and limb darkening parameters. EXOFASTv2 uses several different methods to robustly constrain the stellar parameters. In our calculations we set both NOMIST and TORRES keywords to be able to recover the behavior of the original EXOFAST, using the empirical Torres et al. (2010) relations to derive the stellar mass and radius. EXOFASTv2 is capable of fitting a number of astrometric data sets for a number of planets with RV data sets by scaling the RV and light curve uncertainties.  In this study, we fit our transit light curves along with published RV data. The HATP-36 RV data were obtained by \citet{2012AJ....144...19B} with the Tillinghast Reflector Echelle Spectrograph (TRES) mounted on the 1.5m Fred Lawrence Whipple Observatory (FLWO) telescope;  the HATP-56 RV data were taken at the same facility by \citet{2015AJ....150...85H}, and finally, the WASP-52 RV data were obtained by \citet{2013AandA...549A.134H} using the CORALIE spectrograph (mounted on the 1.2-m Euler-Swiss telescope at La Silla).\\ 
\\
EXOFASTv2 allows the user to supply Gaussian or uniform priors on any fitted or derived parameter.  Priors can simply be listed in a configuration file. In our analysis we have listed a set of stellar and planetary prior parameters which are derived from previous studies in the literature. These prior parameters are $T_{\rm eff}$, $[{\rm Fe/H}]$, $R_*$, $M_*$, linear and quadratic limb darkening coefficients, $R_P$, $M_P$, eccentricity (if available in the literature), period, inclination, and transit impact parameter. We included their uncertainties in our analysis because fixing values is generally not recommended \citep{Eastman2019}, and avoids under estimating the uncertainties of any covariant parameter. \\
\\
To perform the global fits, we used the prior system parameters given by \citet{2012AJ....144...19B} ($R_*$, $M_*$,$T_{\rm eff}$,$[{\rm Fe/H}]$, P, $R_P$ and $M_P$), \citet{2019AJ....157...82W} (eccentricity) for HATP-36, \citet{2015AJ....150...85H} ($R_*$, $M_*$,$T_{\rm eff}$,$[{\rm Fe/H}]$, P, $R_P$ and b) for HATP-56, and \citet{2013AandA...549A.134H} ($R_*$, $M_*$,$T_{\rm eff}$,$[{\rm Fe/H}]$, P) for WASP-52.  The limb-darkening parameters were taken from the NASA Exoplanet Archive\footnote{https://exoplanetarchive.ipac.caltech.edu}. \\
\\
The transit and RV data were independently fitted with EXOFAST and the errors scaled to find the maximum likelihood with lowest $\chi^2$ for every best-fit model. A global fit was performed using both data sets.  The default configuration for the very latest version of EXOFAST (see \citet{Eastman2019}) uses the MIST stellar evolutionary models of \citet{2016ApJS..222....8D}. However, as noted earlier, we opted to use NOMIST and TORRES options during the fitting procedure \citep{2013PASP..125...83E, Eastman2017,Eastman2019} to be consistent with earlier studies. EXOFASTv2 calculates Gelman–Rubin statistic (GR; \citet{Gelman1992}) and the number of independent draws of the underlying posterior probability distribution (Tz; \citet{ford2006}) statistic metrics to judge the convergence. We used default EXOFASTv2 statistics as Tz > 1000 and GR < 1.01 for each parameter to derive the stellar parameters from the global fits. The median values of the posterior distributions for the system parameters, along with the uncertainties (at the $1\sigma$  level), are listed in Tables \ref{tab:syspar_hatp36}, \ref{tab:syspar_hatp56} and \ref{tab:syspar_wasp52}. The posterior distributions obtained for the main stellar and planetary parameters of each system are presented in the (online) Appendix; see Figures \ref{fig:HATP-36b_covar}, \ref{fig:HATP-56b_covar} and \ref{fig:WASP-52b_covar}.   

\subsection{Mid-Transit Times and TTVs}\label{sec:midtransit} 
The possible signature of the existence of a third body (or more bodies) ensues from the systematic variation of the mid-transit times of the system (consisting of a star and an exoplanet). To search for TTVs, calculated and measured mid-transit times are needed. The mid-transit time can be calculated from the following expression: $T_C=T_0 + P_{orb} \times L$,  where $T_0$ is the initial ephemeris time when the cycle is equal to zero,  $P_{orb}$ is orbital period, and $L$ is  the cycle count with respect to the zeroth (or reference) cycle. To extract the observed mid-transit times, we used EXOFASTv2 to separately obtain best-fits for each light curve of our target systems. In addition to our own data, we obtained the published mid-times from the literature. For each system, the observed and the mid-transit times obtained from the literature (along with their uncertainties), are listed in Tables \ref{tab:midtimes_hatp36}, \ref{tab:midtimes_hatp56} and  \ref{tab:midtimes_wasp52} respectively.\\
\\
Using the mid-transit times, we proceeded to obtain the parameters of the linear ephemeris equation by deploying the MCMC affine-invariant ensemble sampler of the emcee package \citep{2010CAMCS...5...65G} implemented by \cite{2013PASP..125..306F}. The priors were taken from calculations based on a linear model using the LMFIT package \citep{2016ascl.soft06014N} by minimizing chi-square. To determine the best-fit parameters of the linear ephemeris model and posterior probability distributions, we used a likelihood function ($\mathcal{L}$) as in \citet{2015MNRAS.448.1118G}:
\begin{equation}
\ln{\mathcal{L}}=
-\frac{1}{2} \sum_{i}^{N} \chi^2
-\sum_{i}^{N}\ln{\sqrt{\sigma_i^2+\sigma_f^2}}
-N\ln{\sqrt{2\pi}}
\end{equation}
where the $\chi^2$ function is given by
\begin{equation}
\chi^2 = \frac{(O-C)_i^2}{\sigma_i^2+\sigma_f^2}
\end{equation}
Here, $(O-C)_i$  denotes the difference between $i$th observed  and calculated ephemeris time, and $\sigma_i$ is the uncertainty in the observed $i$th ephemeris.  This form of $\mathcal{L}$ allows the determination of the free parameter $\sigma_f$ (sometimes called the fractional amount) that scales the raw uncertainties $\sigma_i$ in quadrature \citep{2010CAMCS...5...65G,2015MNRAS.448.1118G}. During sampling the range for the priors of the parameters $T_0, P_{orb}, \sigma_f$ was set to be $> 0$ days.  Then, we performed MCMC for 512 initial walkers (for 30,000 steps), thus providing an effective MCMC chain for the calculation. The resulting 1-D and 2-D posterior probability distributions of the parameters describing the ephemeris (for each system under study) were recorded.
\section{Results and Discussion}
\subsection{System Parameters}\label{sec:sysparameters}
\begin{figure}
	\includegraphics[width=\columnwidth]{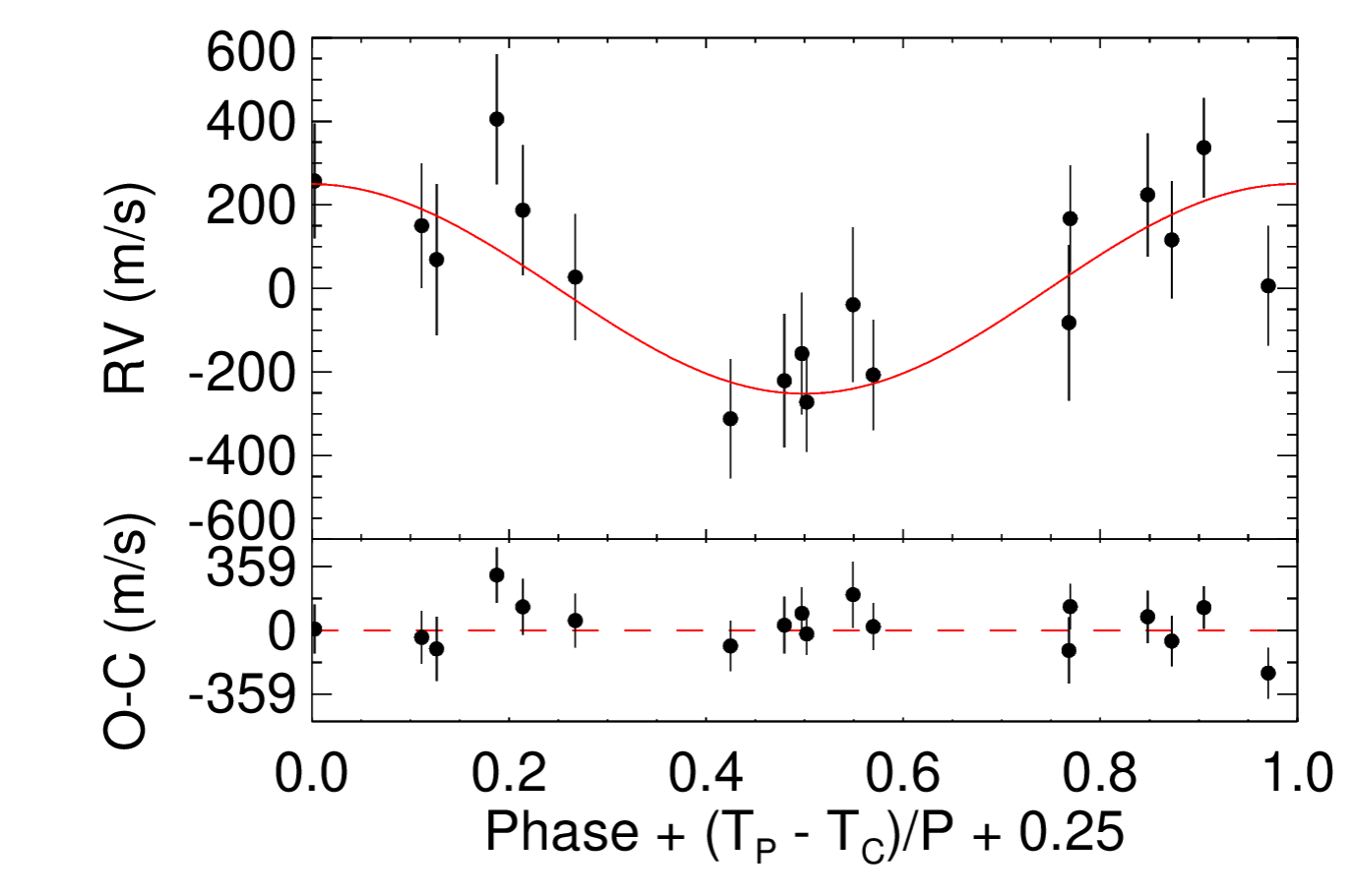}
    \caption{Distribution of radial velocities (RVs) of HATP-56 in the top panel and the residuals in the bottom panel \citep{2015AJ....150...85H}.}
    \label{fig:RV_hatp56}
\end{figure}

\begin{figure}
	\includegraphics[width=\columnwidth]{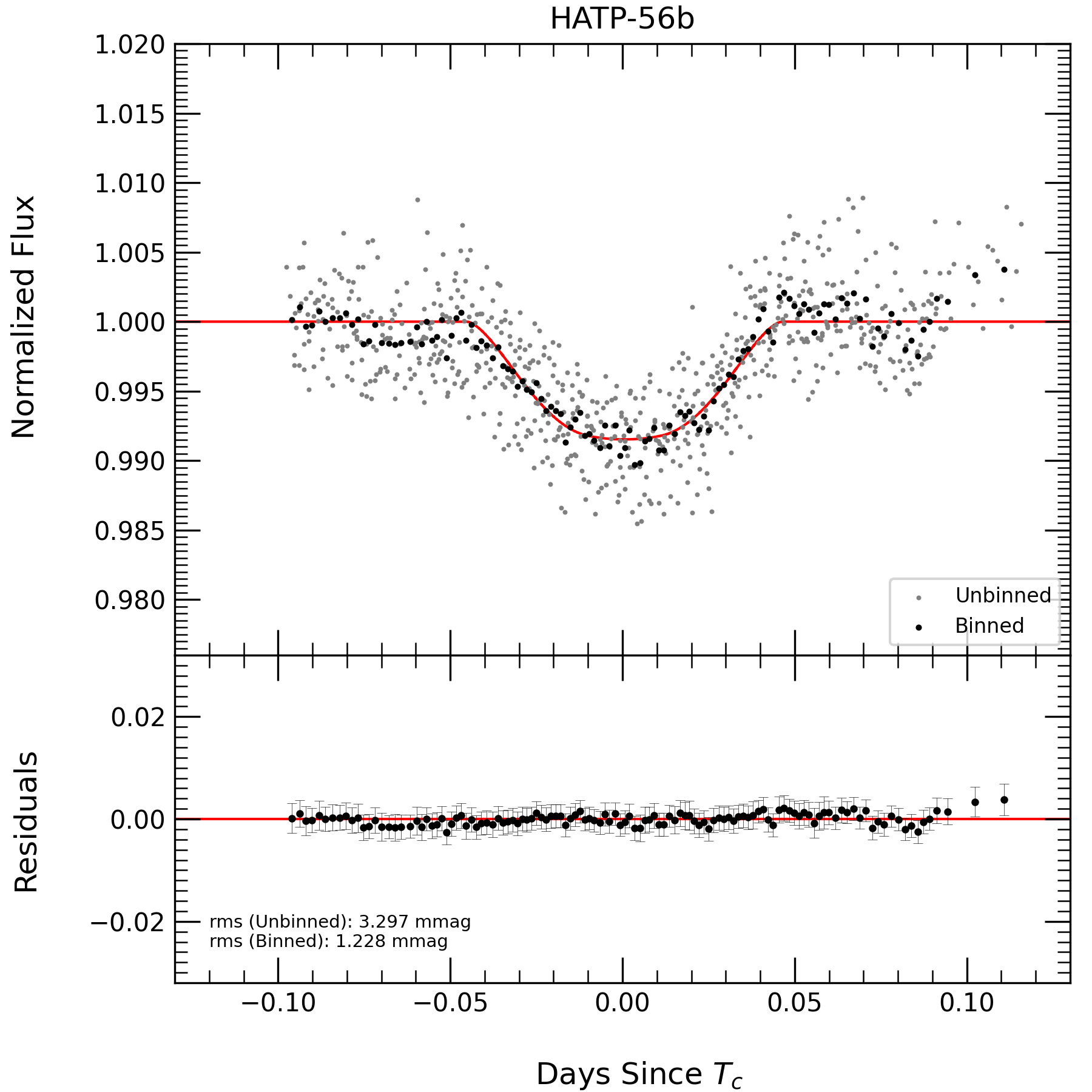}
    \caption{The fitted transit light curves of HATP-56b are shown in the top panel. Unbinned transit light curves are shown as grey points. Binned transit light curves shown as black points. Model residuals for binned transit light curves (for black points) shown in the bottom panel. The RMS of the binned data is  $\sim$ 1.2 mmag and the binning is 147 s.} 
    \label{fig:hatp56b_bin}
\end{figure}

\begin{figure}
	\includegraphics[width=\columnwidth]{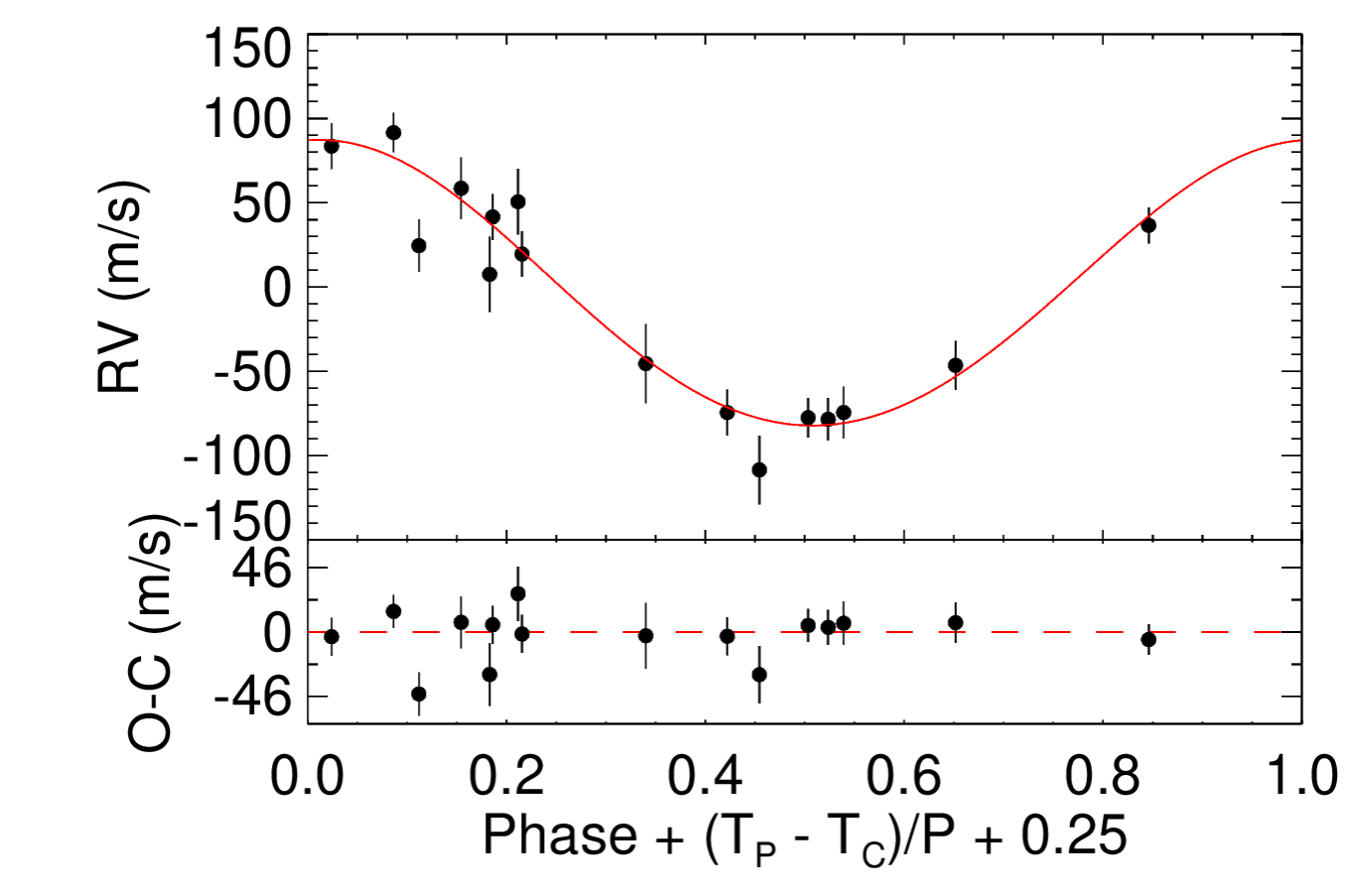}
    \caption{Distribution of radial velocities (RVs) of WASP-52 in the top panel and the residuals in the bottom panel \citep{2013AandA...549A.134H}.}
    \label{fig:RV_wasp52}
\end{figure}

\begin{figure}
	\includegraphics[width=\columnwidth]{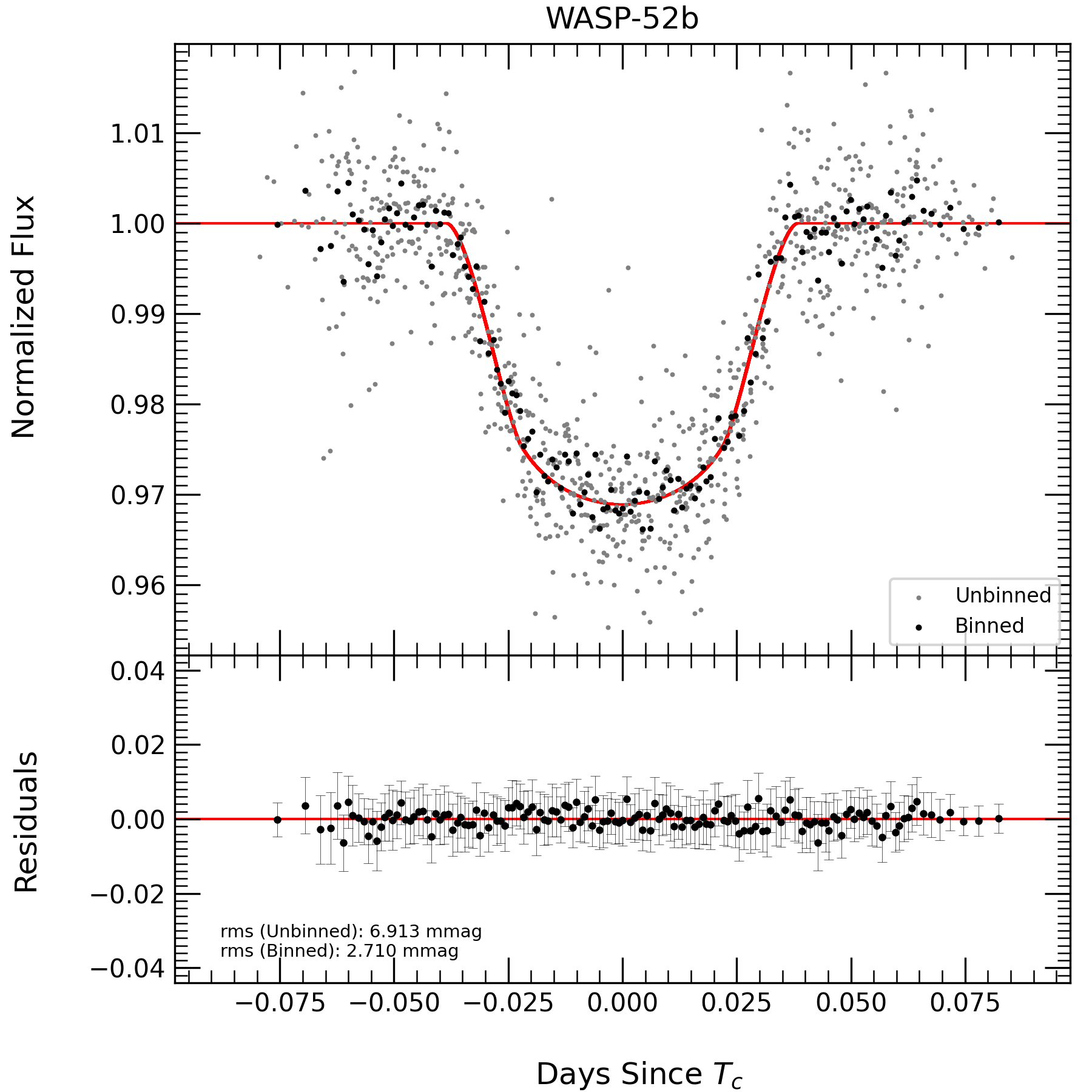}
    \caption{Fitted transit light curves of WASP-52b are shown in the top panel. Unbinned transit light curves are shown as grey points. Binned transit light curves shown as black points. Model residuals for binned transit light curves (for black points) shown in the bottom panel. The RMS of the binned data is  $\sim$ 2.7 mmag and the binning is 90 s.} 
    \label{fig:Wasp52b_bin}
\end{figure}

Based on the global fits with EXOFASTv2, the system parameters we obtained are listed in Tables \ref{tab:syspar_hatp36}, \ref{tab:syspar_hatp56}, and \ref{tab:syspar_wasp52}. The corresponding parameters extracted from previous studies are also listed. Our results are in good agreement with those found in the literature. Our final light curves for HATP-36b, HATP-56b and WASP-52b, along with the transit model fits, are shown in Figures \ref{fig:lcs_hatp36b}, \ref{fig:lcs_hatp56b}, and \ref{fig:lcs_wasp52b} respectively. Figures \ref{fig:RV_hatp36}, \ref{fig:RV_hatp56} and \ref{fig:RV_wasp52}, display the respective fits to the RV data. The global transit fits for the systems are shown in Figures \ref{fig:hatp36b_bin},  \ref{fig:hatp56b_bin} and \ref{fig:Wasp52b_bin} respectively.\\
\\
\subsubsection{HATP-36b}\label{sec:tran_hatp36b}
\begin{figure*}
	\includegraphics[scale=0.8]{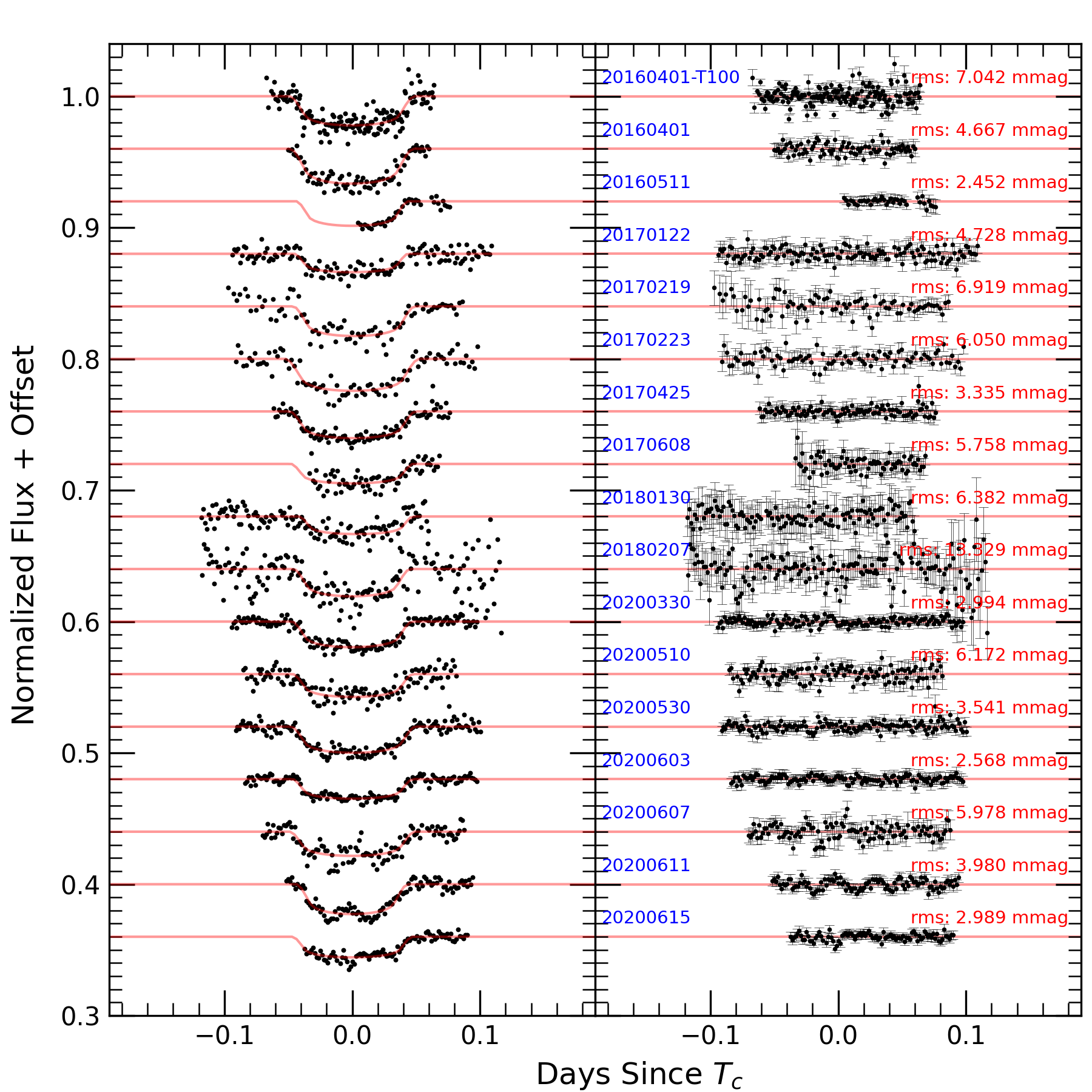}
    \caption{The left panel shows seventeen new transit light curves of HATP-36b along with the transit model fits indicated as red lines. The y-scale is arbitrarily adjusted to provide visual clarity. The right panel displays the residuals along with the dates (in the format: yyyymmdd) and the RMS (in mmag) for each light curve. With the exception of the first lightcurve, which was taken by the TUG facility, the remaining lightcurves were acquired with the ADYU facility.}
    \label{fig:lcs_hatp36b}
\end{figure*}

\begin{figure}
	\includegraphics[width=\columnwidth]{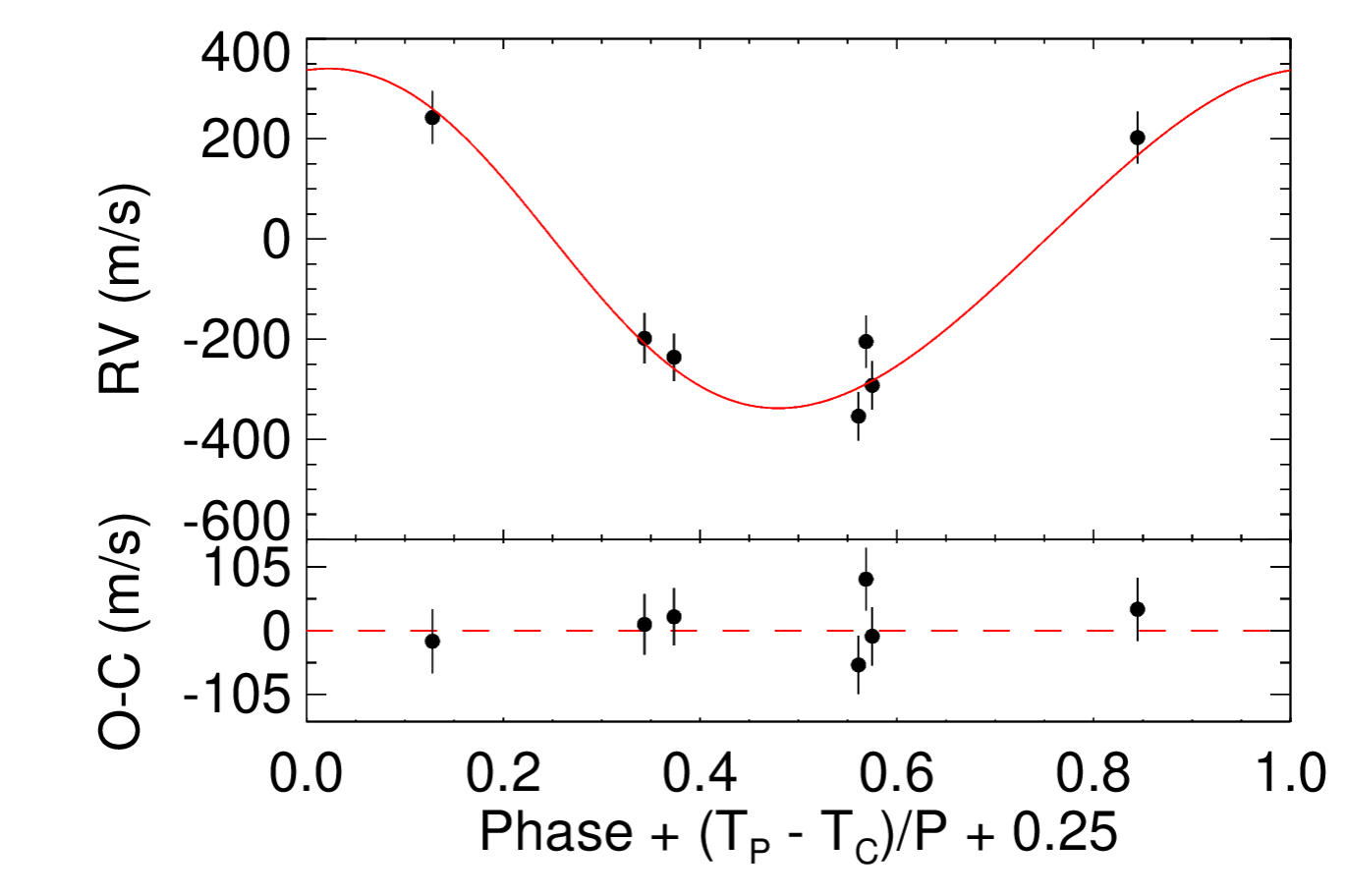}
    \caption{Distribution of the radial velocities (RVs) for HATP-36 in the top panel and the residuals in the bottom panel \citep{2012AJ....144...19B}.}
    \label{fig:RV_hatp36}
\end{figure}

\begin{figure}
	\includegraphics[width=\columnwidth]{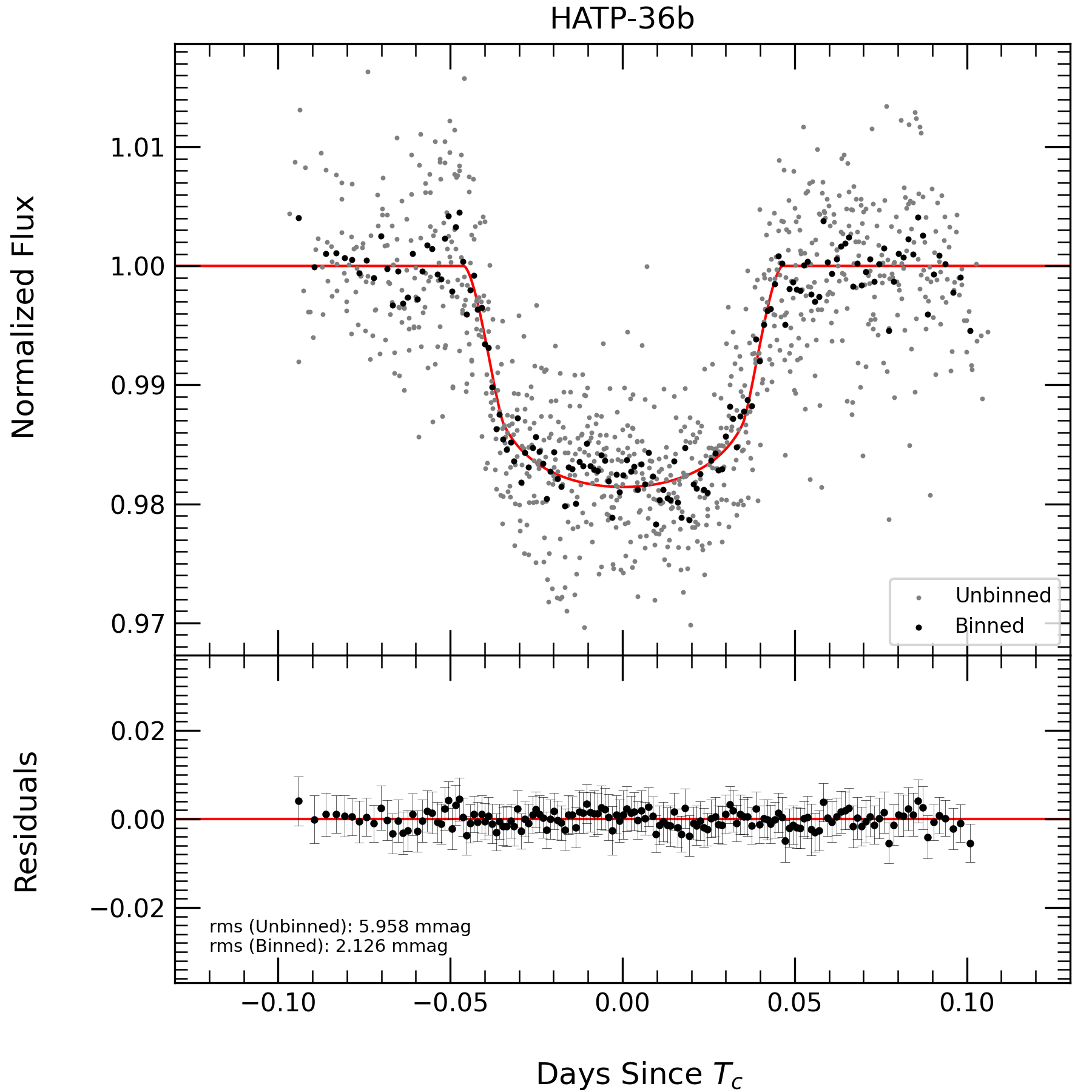}
    \caption{The fitted transit light curves of HATP-36b are shown in the top panel. Unbinned transit light curves are shown as grey points. Binned transit light curves shown as black points. Model residuals for the binned transit light curves (black points) are shown in the bottom panel. The RMS of the binned data is  $\sim$ 2.1 mmag and the binning is 109 s.} 
    \label{fig:hatp36b_bin}
\end{figure}

\begin{figure*}
	\includegraphics[scale=0.8]{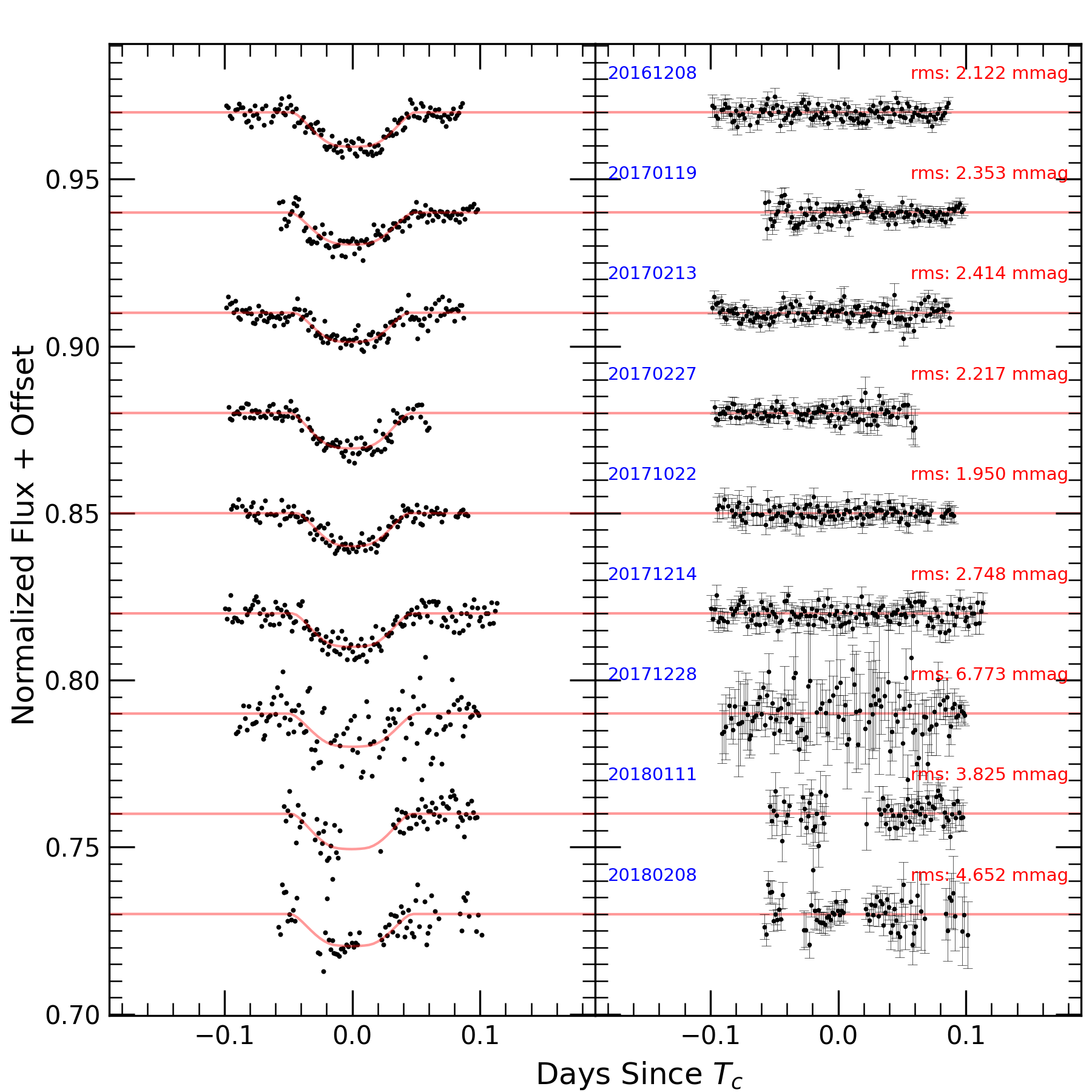}
    \caption{Nine new transit light curves of HATP-56b: All acquired with the ADYU facility. Other description is the same as in Figure \ref{fig:lcs_hatp36b}.}
    \label{fig:lcs_hatp56b}
\end{figure*}

\begin{figure*}
	\includegraphics[scale=0.8]{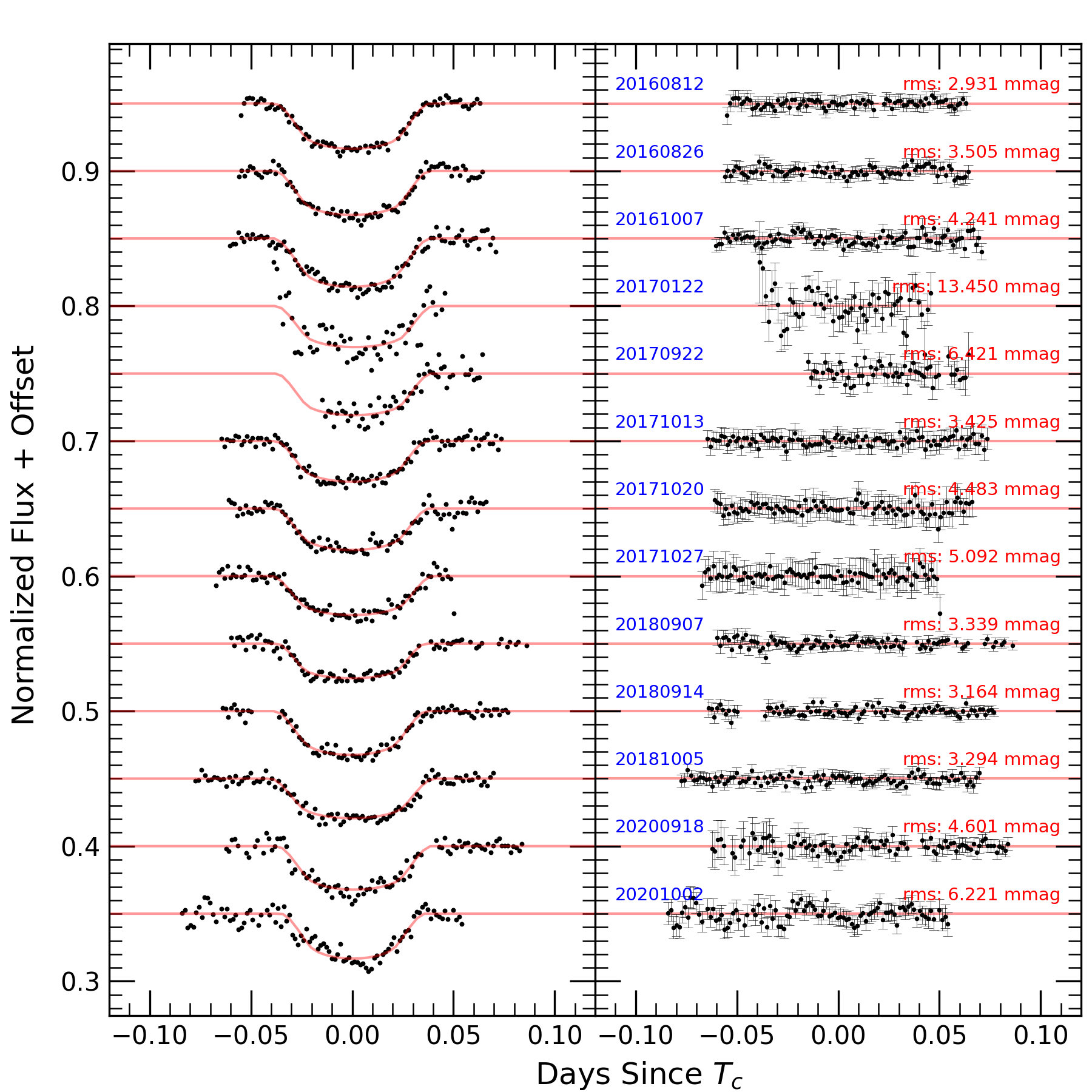}
    \caption{Thirteen new transit light curves of WASP-52b: All acquired with the ADYU facility. The other description is the same as in Figure \ref{fig:lcs_hatp36b}.}
    \label{fig:lcs_wasp52b}
\end{figure*}

Displayed in Figure \ref{fig:lcs_hatp36b} are our 17 transit light curves for HATP-36b along with the global best-fit models. While all the observed light curves are displayed, we only fitted those light curves (a total of 8) that had an RMS of $\leq 6 mmag$ and were complete i.e., possessed both an ingress and egress, as well as, a reasonable baseline. Of course, one wants to deploy as much of the data as one can but a cutoff is still necessary to make sure that suspect and/or highly dispersive data do not dilute the extracted information. We opted for $\leq 6 mmag$ as this provided a good compromise between maximum data and quality. The resulting RMS for the binned transit lightcurve for HAT-P-36b is reasonable i.e., 2.13 mmag and the residuals are also acceptable with no evident structure (see Figure \ref{fig:hatp36b_bin}). The corresponding RV data (including the best-fit) for HATP-36 are shown in Figure \ref{fig:RV_hatp36}.\\
\\
\begin{landscape}
\begin{table}
\caption{System Parameters for HATP-36}
\label{tab:syspar_hatp36}
\begin{tabular}{lccccccc}
\hline\hline 
Parameters & Units & Bakos et al. 2012 & Mancini et al. 2015 & Wang et al. 2019 & This Work \\
\hline
\multicolumn{6}{l}{Stellar parameter:}\\
\hline
$M_{*}$ & Mass $($\(M_\odot\)$)$ & $1.022\pm0.049$ & $1.030 \pm 0.029$ & $1.049_{-0.046}^{+0.048}$ & $1.029\pm0.033$ \\
$R_{*}$ & Radius $($\(R_\odot\)$)$ & $1.096\pm0.056$ & $1.041\pm0.013$ & $1.108_{-0.024}^{+0.025}$ & $1.064\pm0.024$ \\
$L_{*}$ & Luminosity $($\(L_\odot\)$)$ & $1.03\pm0.15$ & $...$ & $1.053_{-0.046}^{+0.048}$ & $0.969_{-0.073}^{+0.078}$ \\
$\rho_{*}$ & Density (cgs) & $...$ & $...$ & $1.089_{-0.024}^{+0.025}$ & $1.205_{-0.070}^{+0.077}$ \\
$log(g_{*})$ & Surface gravity $(cgs)$ & $4.37\pm0.04$ & $4.416\pm0.010$ & $4.37\pm0.04$ & $4.397{\pm0.020}$ \\
$T_{eff}$ & Effective temperature $(K)$ & $5560\pm100$ & $5620\pm40$ & $5560\pm100$ & $5553{\pm 84}$  \\
$[Fe/H]$ & Metallicity & $0.26\pm0.10$ & $0.25\pm0.09$ & $0.26\pm0.10$ & $0.24{\pm0.09}$  \\
\hline
\multicolumn{6}{l}{Planetary Parameters:}\\
\hline
$e$  & Eccentricity & $0.063\pm0.032$ & $...$ & $0.063_{-0.023}^{+0.021}$ & $0.051\pm{0.025}$ \\
$P$     & Period $(days)$ & $1.327347\pm0.000003$ & $1.32734683\pm0.00000048$ & $1.32734660\pm0.00000033$ & $1.32734373\pm0.00000042$ \\
$a$   & Semimajor axis $(au)$ & $0.0238\pm0.0004$ & $0.02388\pm0.00022$ & $0.02402_{-0.00035}^{+0.00036}$ & $0.02388\pm{0.00026}$ \\
$M_p$    & Mass $(M_J)$ & $1.832\pm0.099$ & $1.852\pm0.088$ & $1.925_{-0.081}^{+0.085}$ & $1.871\pm{0.058}$ \\
$R_p$    & Radius $(R_J )$ & $1.264\pm0.071$ & $1.304\pm0.021$ & $1.357_{-0.034}^{+0.035}$ & $1.285\pm0.03$ \\
$\rho_p$    & Density ($\rho_J)$ & $1.12\pm0.19$ & $...$ & $0.955_{-0.054}^{+0.057}$  & $1.093_{-0.075}^{+0.080}$ \\
$\log(g_p)$    & Surface gravity $(cgs)$ & $3.45\pm0.05$ & $...$ &  $3.413\pm0.017$ & $3.448\pm0.022$ \\
$T_{eq}$    & Equilibrium Temperature $(K)$ & $1823\pm55$ & $1788\pm15$ & $1820.2_{-6.8}^{+6.7}$ & $1787\pm32$ \\
\hline
\multicolumn{6}{l}{Primary Transit Parameters:}\\
\hline
$R_P/R_*$    & Radius of planet in stellar radii & $0.1186\pm0.0012$ & $...$ & $0.1260\pm0.0011$  &$0.1242\pm0.0015$ \\
$a/R_*$ & Semi-major axis in stellar radii & $4.66\pm0.22$ & $...$ & ${4.665}_{- 0.034}^{+ 0.035}$ &${4.826}_{- 0.096}^{+ 0.100}$ \\
$u_1$ $(R$ $band)$ & linear limb-darkening coeff & $...$ & $...$ & $0.424$  &${0.420}{\pm 0.010}$ \\
$u_2$ $(R$ $band)$   & quadratic limb-darkening coeff & $...$ & $...$ & $0.250$  &${0.260}\pm0.010$ \\
$i$    & Inclination $(degrees)$ & $86.0\pm1.3$ & $ 85.86\pm0.21$ & $ {85.19}_{- 0.55}^{+ 0.72}$ &$85.60\pm{0.10}$ \\
$b$    & Impact parameter & ${0.312}_{- 0.105}^{+ 0.078}$ & $...$ & ${0.373}_{- 0.062}^{+ 0.048}$  &${0.327}\pm{0.016}$ \\
$T_{0}$ & BJD & $2455565.18144\pm{0.00020}$ & $2455565.18165\pm{0.00037}$ & $2456698.735910\pm{0.000149}$ & $2457402.229738\pm{0.000130}$ \\
\hline
\end{tabular}
\end{table}
\end{landscape}
\begin{figure}
	\includegraphics[width=\columnwidth]{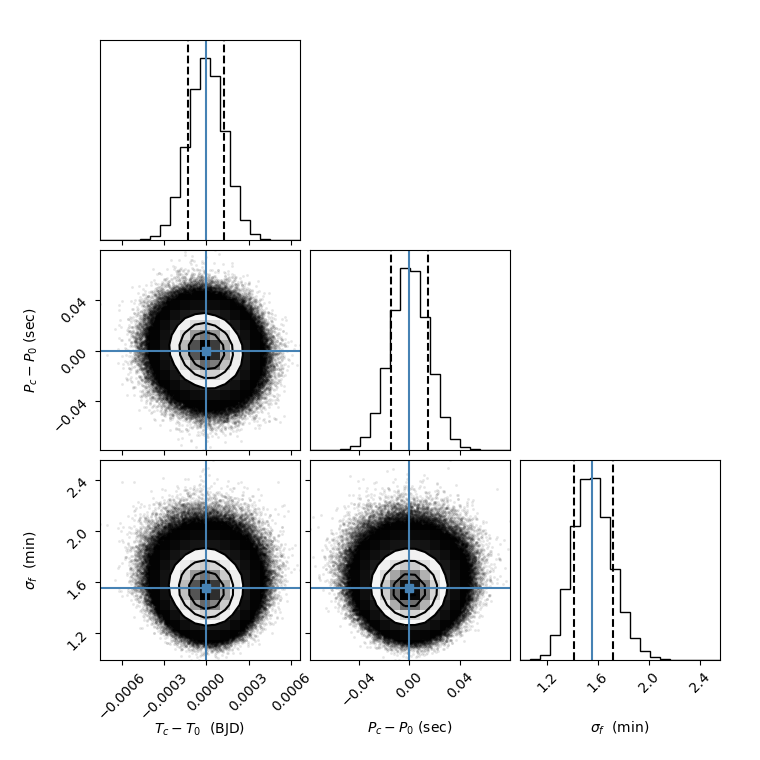}
    \caption{A corner plot showing 1-D and 2-D posterior probability distributions for the parameters of the linear ephemeris for HATP-36b.}
    \label{fig:hatp36b_corner}
\end{figure}
\begin{table*}
\caption{System Parameters for HATP-56}
\label{tab:syspar_hatp56}
\begin{tabular}{lcccc}
\hline\hline 
parameters & Units & Huang et al. 2015 & This work \\ 
\hline
\multicolumn{4}{l}{Stellar parameter:}\\
\hline
$M_{*}$  & Mass $($\(M_\odot\)$)$ & $1.296\pm0.036$ & $1.315\pm0.029$\\ 
$R_{*}$  & Radius $($\(R_\odot\)$)$ & $1.428\pm0.030$ & $1.430\pm0.019$\\ 
$L_{*}$  & Luminosity $($\(L_\odot\)$)$ & $3.390\pm0.19$ & $3.420\pm{0.14}$\\ 
$\rho_{*}$  & Density (cgs) & $0.627\pm0.033$ & $0.633\pm{0.025}$\\ 
$log(g_{*})$ & Surface gravity $(cgs)$ & $4.240\pm0.015$ & $4.246\pm0.012$\\ 
$T_{eff}$ & Effective temperature $(K)$ & $6566\pm50$ &  $6563\pm{49}$\\ 
$[Fe/H]$ & Metallicity  & $-0.077\pm0.080$ &  $-0.068 {\pm0.078}$\\ 
\hline 
\multicolumn{4}{l}{Planetary Parameters:}\\ 
\hline 
$e$ & Eccentricity & $ <0.246 $ & $0.028^{+0.042}_{-0.020}$&\\ 
$P$ & Period $(days)$ & $2.7908327\pm0.0000047$ &  $2.7908278\pm{0.0000019}$\\ 
$a$ & Semimajor axis $(au)$ & $0.04230\pm0.00039$ & $0.04252\pm0.00031$\\ 
$M_p$ & Mass $(M_J )$ & $2.18\pm0.25$ &  $2.11 {\pm 0.21}$\\ 
$R_p$ & Radius $(R_J )$ & $1.466\pm0.040$ &  $1.387\pm0.029$\\ 
$\rho_p$ & Density ($\rho_J)$ & $0.86\pm0.12$ & $0.98{\pm 0.12}$\\ 
$\log(g_p)$ & Surface gravity & $3.402\pm0.055$ & $3.435{\pm0.047}$\\ 
$T_{eq}$ & Equilibrium Temperature $(K)$ & $1840\pm21$ & $1835\pm18$\\ 
\hline 
\multicolumn{4}{l}{Primary Transit Parameters:}\\
\hline 
$R_P/R_*$ & Radius of planet in stellar radii & $0.10540\pm0.00086$ &  $0.0997\pm0.0019$\\ 
$a/R_*$ & Semi-major axis in stellar radii & $6.37\pm0.11$ &  $6.391{\pm 0.084}$\\ 
$u_1$ $(R$ $band)$ & linear limb-darkening coeff & $...$ &  $0.319{\pm0.049}$\\ 
$u_2$ $(R$ $band)$ & quadratic limb-darkening coeff & $...$ & $0.371{\pm0.049}$\\ 
$i$ & Inclination $(degrees)$ & $82.13\pm0.18$ & $82.02\pm0.15$\\ 
$b$ & Impact parameter & $0.8725^{+0.0044}_{-0.0060}$ & $0.8806{\pm 0.0055}$\\ 
$T_{0}$ & BJD &  $2456553.61645\pm0.00042$  &  $2457731.345459\pm0.000775$  \\
\hline
\end{tabular}
\end{table*}
\begin{table*}
\caption{System Parameters for WASP-52}
\label{tab:syspar_wasp52}
\begin{tabular}{lccccc}
\hline\hline 
parameters & Units & Hebrard et al 2013 & Mancini et al. 2017 & This work \\ 
\hline
\multicolumn{4}{l}{Stellar parameter:}\\
\hline
$M_{*}$ & Mass $($\(M_\odot\)$)$ & $0.87\pm0.03$ & ${0.804\pm0.050}$ & $0.857\pm0.020$\\ 
$R_{*}$ & Radius $($\(R_\odot\)$)$ & $0.79\pm0.02$ & ${0.786\pm0.016}$ & $0.791\pm0.008$\\ 
$L_{*}$ & Luminosity $($\(L_\odot\)$)$ & $...$  &  $...$  & ${0.363}{\pm0.025}$\\ 
$\rho_{*}$ & Density $($\(\rho_\odot\)$)$ & $1.76\pm0.08$ & $1.65\pm0.02$  & ${1.73}{\pm 0.06}$\\ 
$log(g_{*})$ & Surface gravity $(cgs)$ & $4.582\pm0.014$ & ${4.553\pm0.010}$ & ${4.575}{\pm0.008}$\\ 
$T_{eff}$ & Effective temperature $(K)$ & $5000\pm100$   & $...$  & $5035\pm82$\\ 
$[Fe/H]$ & Metallicity  & $0.03\pm0.12$  & $...$  & $0.08\pm0.11$\\ 
\hline  
\multicolumn{4}{l}{Planetary Parameters:}\\ 
\hline  
$e$ & Eccentricity & ${0(fixed)}$  & $...$  & ${0.026}_{-0.016}^{+0.019}$\\ 
$P$ & Period $(days)$ & $1.7497798\pm0.0000012$ & $1.74978119\pm0.00000052$ & ${1.74978111}{\pm0.00000093}$\\ 
$a$ & Semimajor axis $(au)$ & $0.0272\pm0.0003$ & ${0.02643\pm0.00055}$ & ${0.027\pm0.001}$\\ 
$M_p$ & Mass $(M_J )$ & $0.46\pm0.02$ & ${0.434\pm0.024}$ & ${0.443\pm0.012}$\\ 
$R_p$ & Radius $(R_J )$ & $1.27\pm0.03$ & ${1.253\pm0.027}$ & ${1.281\pm0.011}$\\ 
$\rho_p$ & Density ($\rho_J)$ & $0.22\pm0.02$ & ${0.2061\pm0.0091}$ & ${0.2614}{\pm0.0083}$\\ 
$\log(g_p)$ & Surface gravity $(cgs)$ & $2.81\pm0.03$ & $...$  & ${2.83\pm0.01}$\\ 
$T_{eq}$ & Equilibrium Temperature $(K)$ & $1315\pm35$ & $1315\pm26$ & ${1314}{\pm 22}$\\ 
\hline  
\multicolumn{4}{l}{Primary Transit Parameters:}\\ 
\hline  
$R_P/R_*$ & Radius of planet in stellar radii & $0.1646\pm0.0012$ & $0.16378\pm0.0005$ & ${0.1665}{\pm0.0014}$\\ 
$a/R_*$ & Semi-major axis in stellar radii & $7.38\pm0.02$ & $...$  & ${7.34}{\pm0.06}$\\ 
$u_1$ $(R$ $band)$ & linear limb-darkening coeff. &$...$   & $...$  & ${0.539\pm0.034}$\\ 
$u_2$ $(R$ $band)$ & quadratic limb-darkening coeff & $...$  & $...$  & ${0.192\pm0.035}$\\ 
$i$ & Inclination $(degrees)$ & $85.35\pm0.20$ & $85.15\pm0.06$ & ${85.35\pm0.01}$\\ 
$b$ & Impact parameter & $0.60\pm0.02$ & $...$  & ${0.59}{\pm 0.01}$\\ 
$T_{0}$ & BJD &  $2455793.682175\pm0.00009$  &    $2456862.79776\pm0.00016$  &  $2457669.447361\pm0.000210$ \\
\hline
\end{tabular}
\end{table*}

\begin{figure}
	\includegraphics[width=\columnwidth]{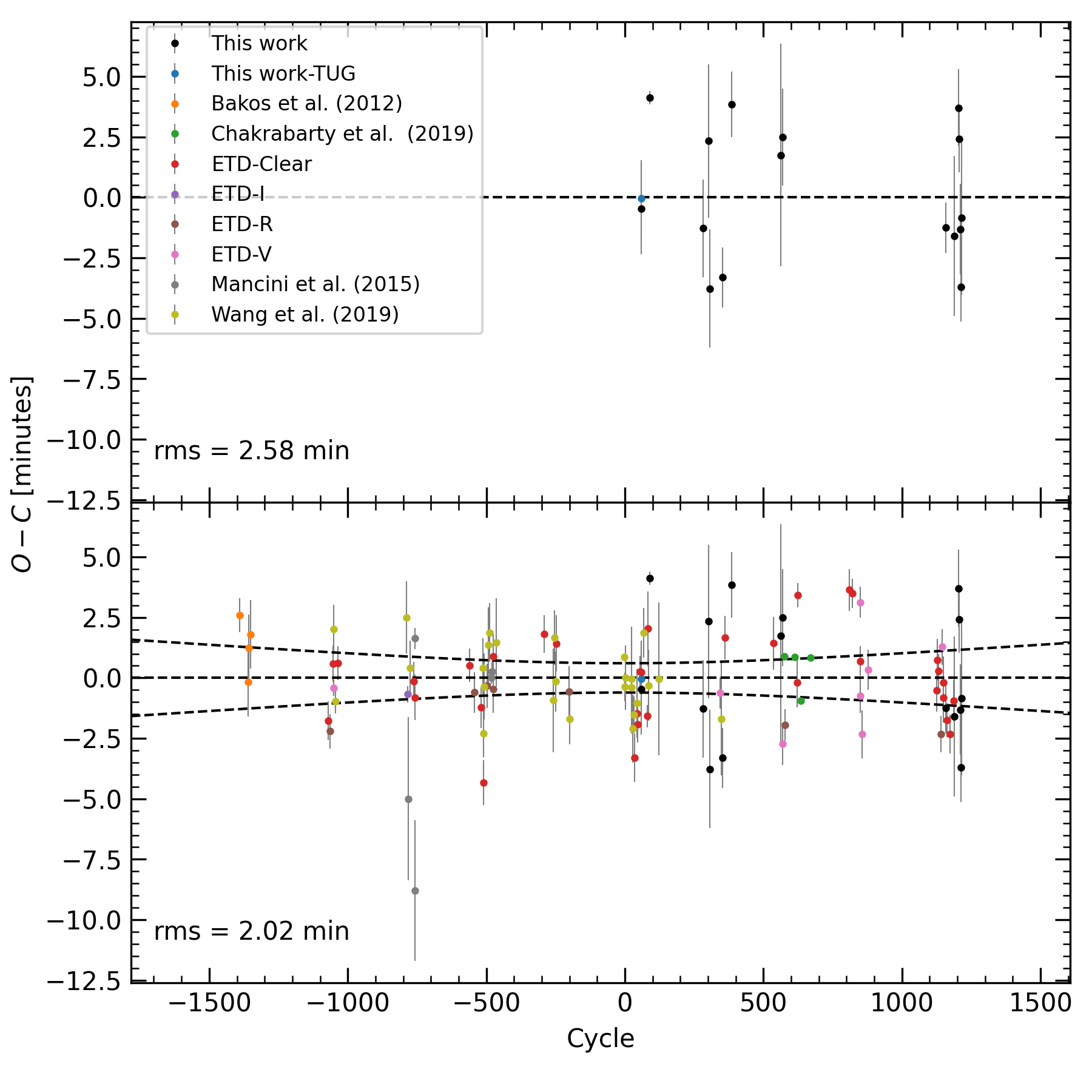}
    \caption{TTVs for HATP-36b.  Our results (upper panel: black and blue points). Our results plus data taken from the literature (lower panel: colored points). Dashed (curved) lines indicate the $\pm$3$ \sigma$ confidence levels. The RMS values for the data sets are  $\sim$ 2.6 min and $\sim$ 2.0 min respectively.}
    \label{fig:OCs_hatp36b}
\end{figure}

\noindent To perform the TTV analysis for HATP-36b, precise mid-transit times were determined from each transit light curve, which was separately modeled by EXOFASTv2. The mid-transit time obtained for each light curve, as well as ephemeris times collected from the literature \citep{2012AJ....144...19B, 2015AandA...579A.136M, 2019AJ....158...39C, 2019AJ....157...82W} are listed in Table \ref{tab:midtimes_hatp36}. In addition, we also collected mid-transit times from the Exoplanet Transit Database (ETD) \citep{2010NewA...15..297P}. In the ETD, the observational data are rated by a quality index in the range 1 - 5, where the quality increases with decreasing index. Thus, we only chose the data indicated by indices 1 or 2.  We converted the mid-transit times of ETD in units of Heliocentric Julian Date (HJD) to BJD through the web-gui tool\footnote{http://astroutils.astronomy.ohio-state.edu/time/} \citep{2010PASP..122..935E}. We also note that the ETD data contain observations with different filters (I, V, and R) indicated in the figures as colored points and are included in our TTV calculations. All 88 published mid-transit data cover the time span 2010 - 2020.\\ 
\\
 We fitted the extracted mid-transit times, together with those taken from the literature, with a linear function of the form; 
\begin{equation}
T_C=T_0 + P_{orb} \times L
\end{equation}
\noindent
Here, the mid transit time when epoch is equal to zero $L=0$ is $T_0= {2457402.229738}{\pm0.000130}$~[BJD$_{TDB}$], and the orbital period is $P_{orb}= {1.32734733}{\pm0.00000017}$ days and $\sigma_f$ = 1.56$_{-0.14}^{+0.16}$ mins. For completeness, 1-D and 2-D posterior probability distributions for the parameters of the updated linear ephemeris are shown as a corner plot in Figure \ref{fig:hatp36b_corner}.\\ 
\\
 From all mid-times, the \textit {observed - calculated} (O-C) mid transit times were determined, and the O-C diagram was constructed; the plot is shown in Figure \ref{fig:OCs_hatp36b}. Our results are shown in the upper panel (RMS $\sim 2.6 min$); the lower panel displays our results overlaid with O-C data taken from the literature along with dashed lines indicating the $\pm$3$ \sigma$ confidence levels. With the possible exception of three potential outliers, the majority of the points (from this work) are consistent within the 3$ \sigma$ limit. As noted in the introduction, our prime motivation for monitoring HATP-36b was to follow-up on the possibility of observing stellar activity related to the surface of the host star as suggested by the work of \citet{2015AandA...579A.136M}. Our combined results, based on the analysis of 17 new light curves spanning a time period of approximately four years, suggest the absence of such stellar activity. However, we do note that the dispersion in our data is relatively large so it is possible the effect is masked. Moreover, as is likely, the stellar activity, if it exists (and assuming it to be similar to that observed on the surface of the Sun (\citet{2003ApJ...591..406B} and references therein)), it presumably has a finite lifetime over which it persists with significant magnitude and is negligible at other times thus suggesting another possibility for its absence in our data. In principle, this second scenario is testable with an extended observation campaign.
\\
\begin{figure}
	\includegraphics[width=\columnwidth]{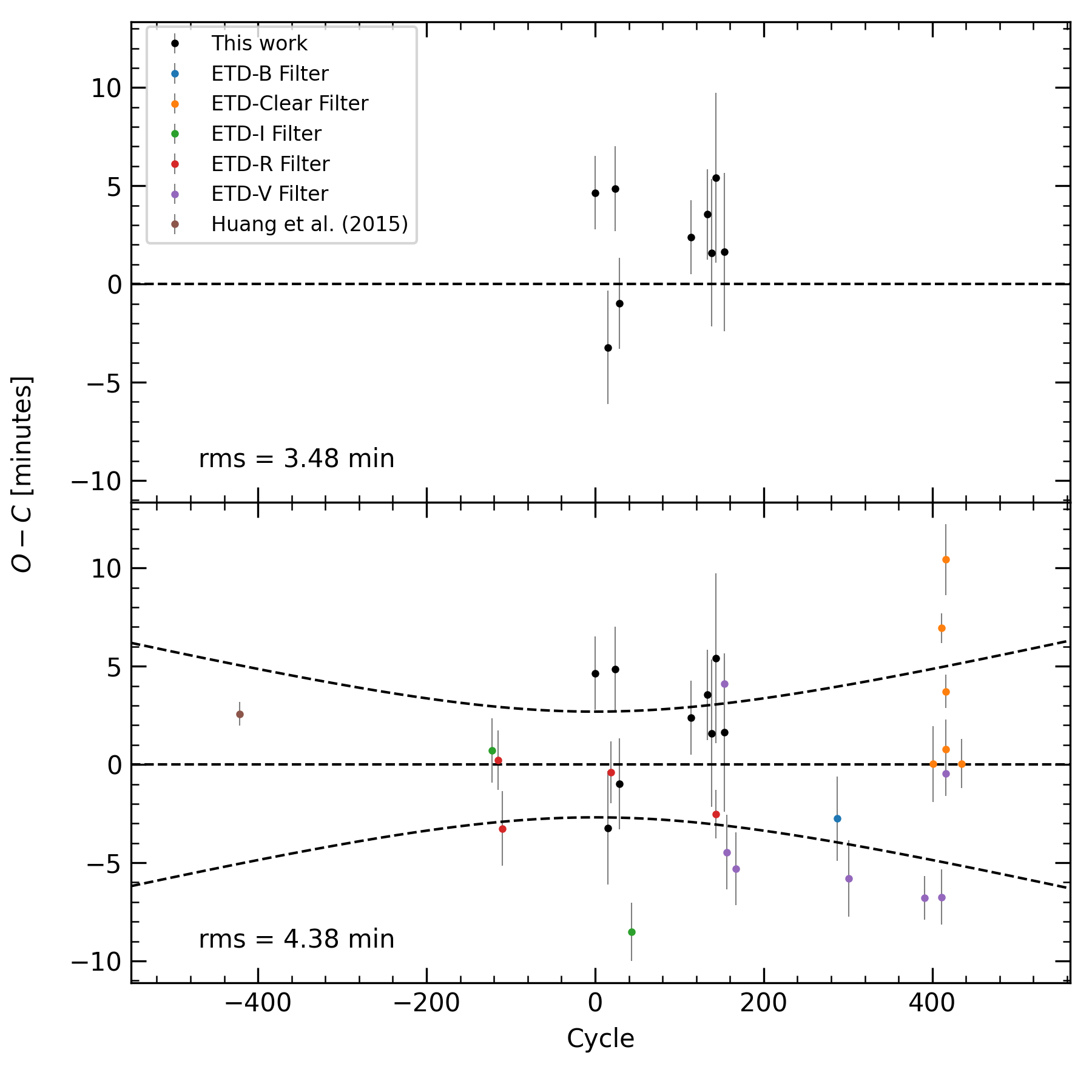}
    \caption{TTVs for HATP-56b. The description is the same as in Figure \ref{fig:OCs_hatp36b} The RMS values for the data sets are  $\sim$ 3.5 min and $\sim$ 4.4 min respectively.}
    \label{fig:OCs_hatp56b}
\end{figure}
\begin{figure}
	\includegraphics[width=\columnwidth]{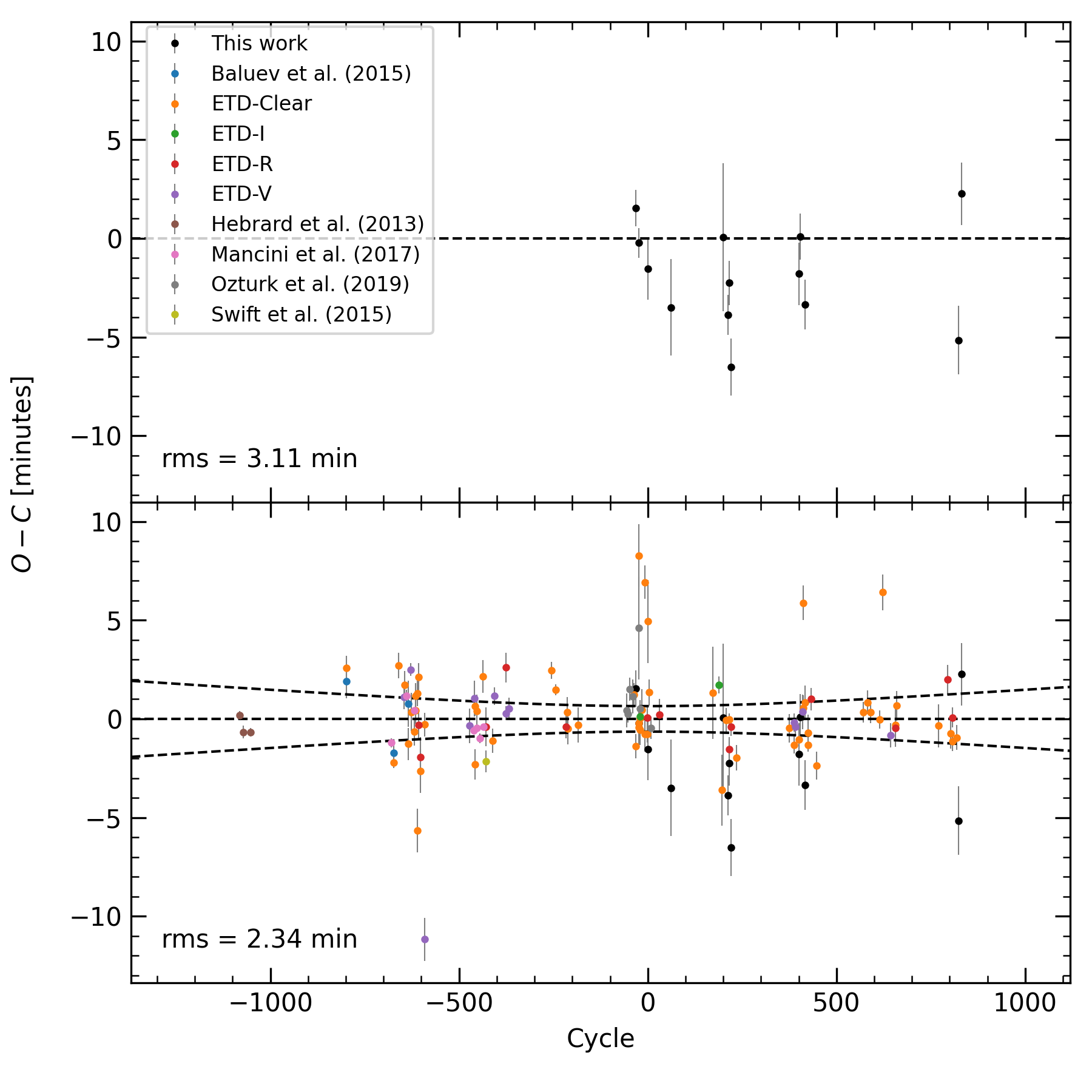}
    \caption{TTVs for WASP-52b. The description is the same as in Figure \ref{fig:OCs_hatp36b}. The RMS values for the data sets are  $\sim$ 3.1 min and $\sim$ 2.3 min respectively.}
    \label{fig:OCs_wasp52b}
\end{figure}
\\
\subsubsection{HATP-56b}\label{sec:tran_hatp56b}
\begin{figure}
	\includegraphics[width=\columnwidth]{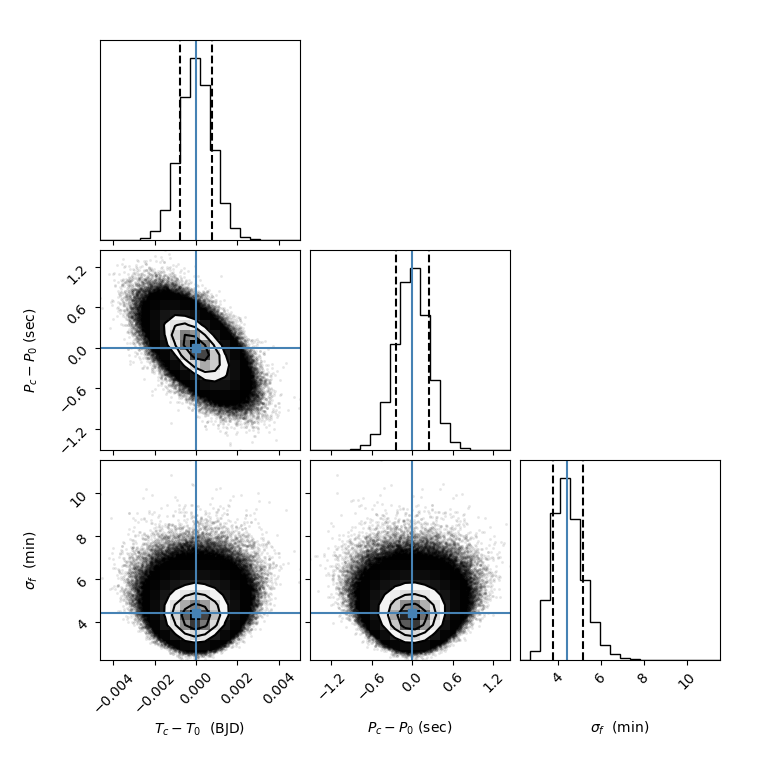}
    \caption{A corner plot showing 1-D and 2-D posterior probability distributions for the parameters of the linear ephemeris for HATP-56b.}
    \label{fig:hatp56b_corner-june2}
\end{figure}

In Figures \ref{fig:lcs_hatp56b} and \ref{fig:RV_hatp56}, we display the 9 transit light curves of HATP-56b (although we fitted only a total of 6 of these with an RMS $\leq 3 mmag$) along with global fits and the RV data, respectively. We found only 22 published mid-transit times covering the time period between 2016-2020. Our additional 9 light curves, obtained in the period 2016-2018, has significantly increased the observations for this system. Our observations and the mid-transit times collected from the literature (\citet{2010NewA...15..297P}, \citet{2015AJ....150...85H}) are listed in Table \ref{tab:midtimes_hatp56}. From the ETD data, we chose the data labeled with indices 1, 2 and 3. As stellar activity has been reported for this system, we opted to include as many reasonable data points as possible, including light curves with indices = 3, so that we could assess the robustness of the claims of intrinsic variability. \\  
\\
From our best-fits to the mid-transit times we obtained updated $T_0$, $P_{orb}$, and $\sigma_f$ as: $T_0= {2457731.345473}{\pm 0.000781}$~[BJD$_{TDB}$],  $P_{orb}= {2.79083132} {\pm 0.00000285}$ and $\sigma_f$ = 4.42$_{-0.63}^{+0.79}$ mins respectively. Using this information, we constructed the O-C diagram that is shown in Figure \ref{fig:OCs_hatp56b}. Our results are shown by themselves in the upper panel (RMS $\sim 3.5 min$); the lower panel of the figure shows our results (black points) overlaid with additional data taken from the literature (colored points) along with dashed lines indicating the $\pm$3$ \sigma$ confidence levels. The majority of the data points are consistent within the 3$ \sigma$ limit although, once again, there appears to be at least one point (from this work) as a possible outlier. The 1-D and 2-D posterior probability distributions for the parameters of the updated linear ephemeris are shown in Figure \ref{fig:hatp56b_corner-june2}.\\
\begin{figure}
	\includegraphics[width=\columnwidth]{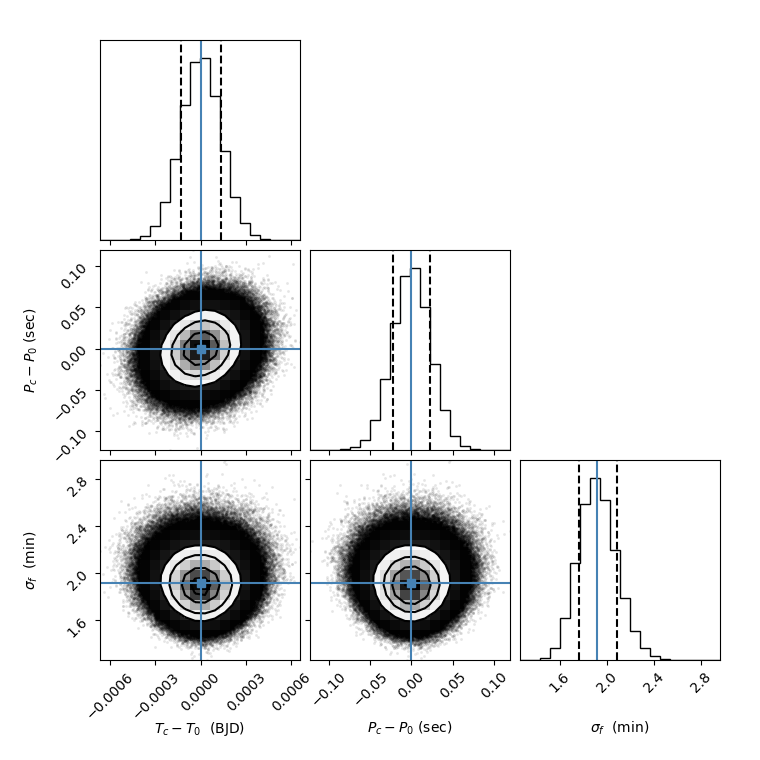}
    \caption{A corner plot showing 1-D and 2-D posterior probability distributions for the parameters of the linear ephemeris for WASP-52b.}
    \label{fig:wasp52b_corner-april26}
\end{figure}
\subsubsection{WASP-52b}\label{sec:tran_wasp52b}
\begin{figure}
	\includegraphics[width=\columnwidth]{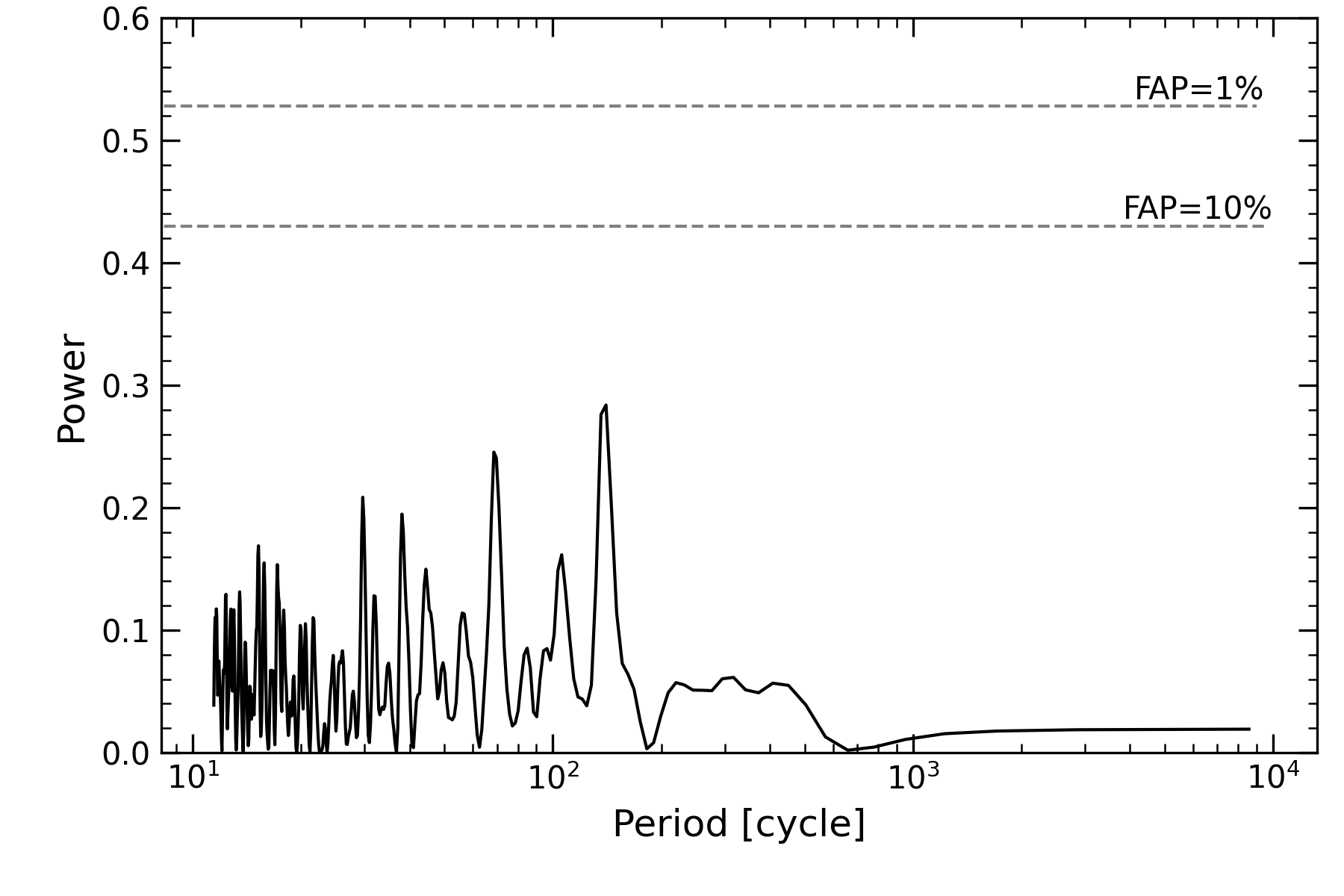}
    \caption{The Lomb-Scargle periodogram for the TTVs of HATP-56b. The false alarm probability (FAP) levels of 1$\%$ and 10$\%$ are shown as dotted horizontal lines. No significant power is detected in the observed cycle range.}
    \label{fig:LS_hatp56b}
\end{figure}

\begin{figure}
	\includegraphics[width=\columnwidth]{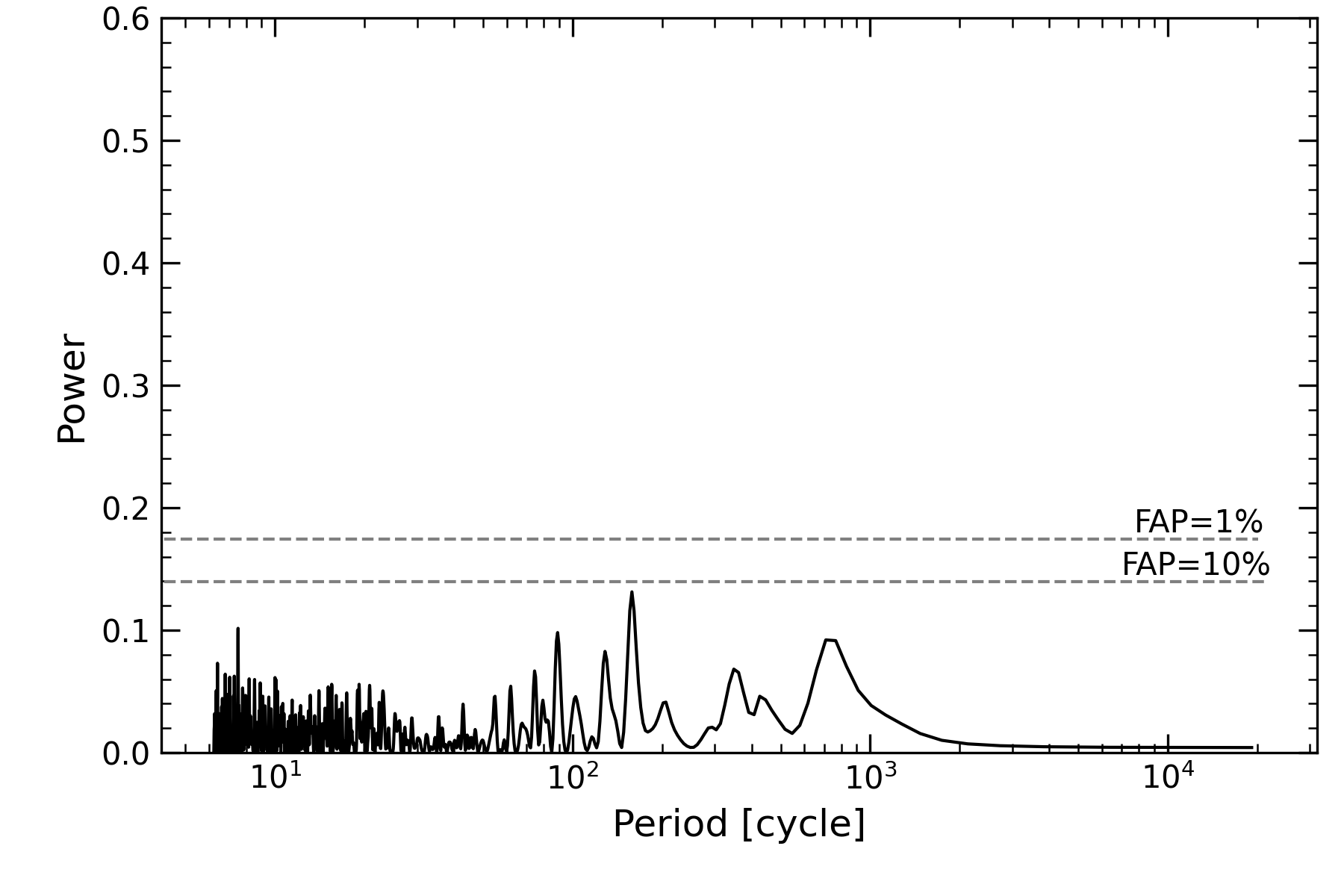}
    \caption{The Lomb-Scargle periodogram for the TTVs of WASP-52b. The other description is the same as in Figure \ref{fig:LS_hatp56b}.}
    \label{fig:LS_wasp52b}
\end{figure}

\begin{figure}
	\includegraphics[width=\columnwidth]{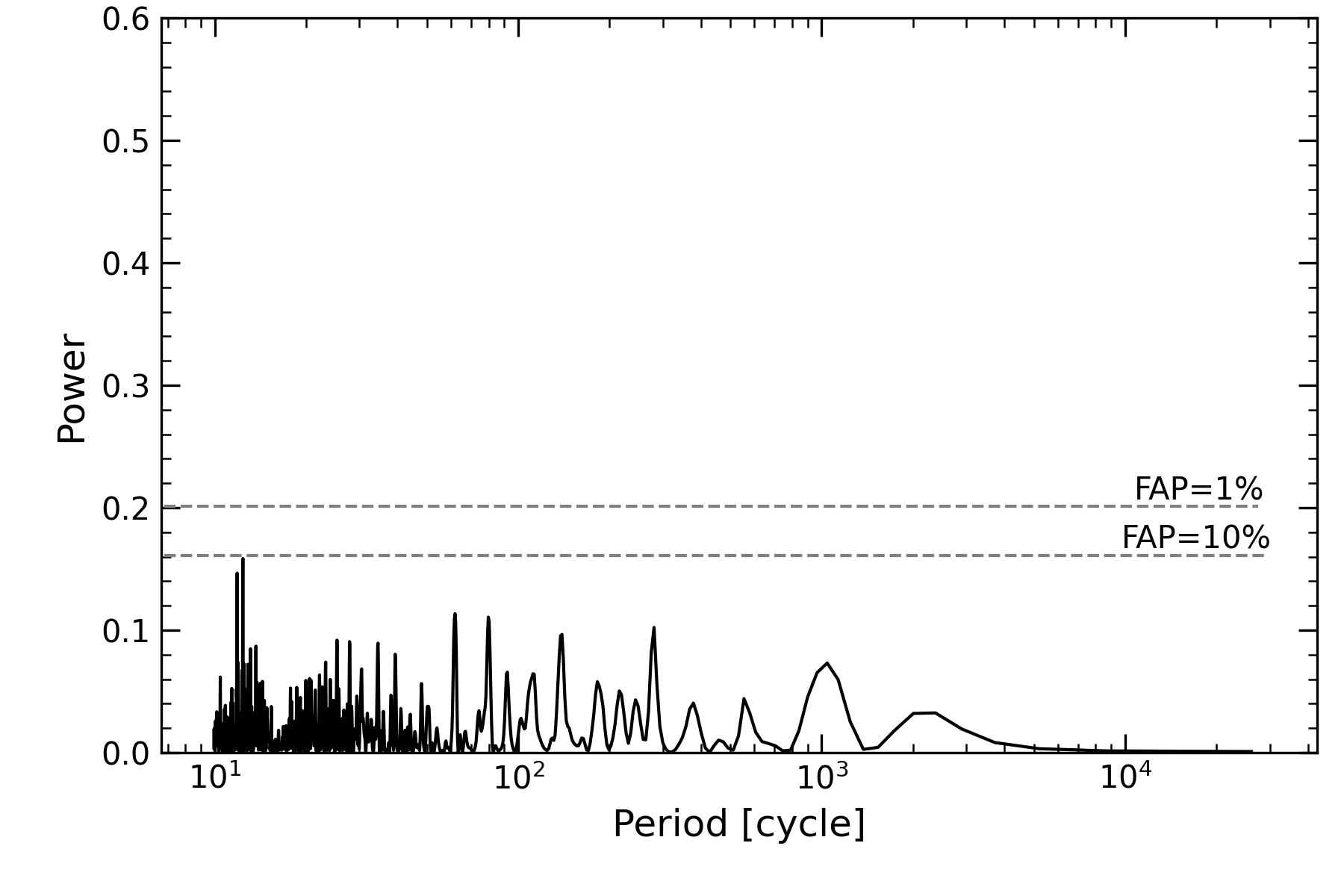}
    \caption{The Lomb-Scargle periodogram for the TTVs of HATP-36b. A narrow,  relatively weak low-power structure at $\sim$ cycle 12.4 with FAP levels of $\sim$ 10$\%$ is noticeable.} 
    \label{fig:LS_hatp36b}
\end{figure}

The 13 newly observed transits along with the global best-fits for WASP-52b and the corresponding RV data are plotted in Figures \ref{fig:lcs_wasp52b} and \ref{fig:RV_wasp52}, respectively. Out of the 13 light curves only 10, with an RMS $\leq 6 mmag$, were included in the global fit. The published data were taken from 5 different studies \citep{2013AandA...549A.134H, 2015MNRAS.450.3101B, 2015JATIS...1b7002S, 2017MNRAS.465..843M, 2019MNRAS.486.2290O} and ETD. We chose the data labeled with indices 1 and 2. The mid-transit times for WASP-52b are listed in Table \ref{tab:midtimes_wasp52}.  Majority of the published light curves are for the period 2010-2016. Our light curves cover the time span between 2016 to 2020, thus extending the overall time span to a total of 10 years. From the best-fitting models for each of our light curves, the mid-transit times were obtained. By fitting a linear function, the initial transit ephemeris, the orbital period and $\sigma_f$ were updated as follows: $T_0 =  {2457669.447361} {\pm0.000210}$~[BJD$_{TDB}$], $P_{orb}={1.74978145} {\pm 0.00000042}$ days, and $\sigma_f$ = 1.92$_{-0.15}^{+0.17}$ respectively. The (O-C) values are given in Table \ref{tab:midtimes_wasp52}, and the corresponding O-C diagram is shown in Figure \ref{fig:OCs_wasp52b}.  The upper panel of the figure displays our results (RMS $\sim 3.1 min$) by themselves whereas the lower panel shows the combined data (our results, black points, plus those taken from the literature as colored points) along with curved dashed lines indicating the $\pm$3$ \sigma$ confidence levels. The majority of the data points are consistent within the 3$ \sigma$ limit although, once again, there appear to be possible outliers (five points from this work). The 1-D and 2-D posterior probability distributions for the parameters of the updated linear ephemeris are shown in Figure \ref{fig:wasp52b_corner-april26}.\\
\\
As a sanity check, we constructed Lomb-Scargle periodograms \citep{1976Ap&SS..39..447L, 1982ApJ...263..835S} for the three systems under investigation (see Figures \ref{fig:LS_hatp56b}, \ref{fig:LS_wasp52b}, and \ref{fig:LS_hatp36b}). The periodograms provide a measure of the distribution of power in the TTVs. The calculation is based on the O-C values and the corresponding epochs. The theoretical values of the false alarm probabilities (FAPs) of 1$\%$ and 10$\%$ are shown as dashed lines. No significant power is seen for HATP-56b and WASP-52b in the observed cycle range thus supporting the absence of structures in the respective O-C diagrams at the 3$ \sigma$ level. We do note the presence of a relatively weak structure in HATP-36b ($\sim$ cycle 12.4 in Figure \ref{fig:LS_hatp36b}). Given the dispersion in the O-C diagram and a FAP of $\sim$10$\%$, the significance of this low-power structure is not clear.
\\
\section{Summary and Conclusions}\label{sec:conc}
We have performed long-term observations of HATP-36b, HATP-56b and WASP-52b using the 0.6 m and 1.0 m telescope facilities at Adiyaman University and TUBITAK National Observatory respectively. We have measured 17 new transit light curves for HATP-36b, 9 new transit light curves for HATP-56b, and 13 new transit light curves for WASP-52b. Using these observations in concert with published data, we have determined an updated linear ephemeris for each system. In addition, we have used these data to search for the existence of additional bodies in these systems by extracting the TTVs. O-C diagrams are constructed for each system. We summarize our main findings as follows:\\
\begin{itemize}
\item The newly extracted planetary system parameters are in very good agreement with those found in previous studies.
\\
\item We have determined a new ephemeris for each system. Based on the extracted periods and mid-transit times, we note that for all of the systems under study, our values agree with the respective published results within 1-2$\sigma$. 
\\
\item An earlier study reported stellar activity on the surface of the host star in the HATP-36 system; we find no evidence of similar activity in the newly acquired 17 transit light curves spanning a time period of approximately 4 years. Indeed, we find no convincing evidence for stellar activity in the three hot-Jupiters we studied. It is possible the effect is masked in our data because of the significant dispersion. It is also possible that the stellar activity was negligible in magnitude during the period of our observations. The o-c diagrams do indicate variation at the 3$\sigma$ confidence level for a number of data points for each of the systems. However, with the level of scatter/dispersion present in the data and the irregularity of the variations it is far from clear that the potential outliers are indicative of stellar activity associated with the surface of the respective host star.
\\
\item We do not find strong evidence for the existence of a third (gravitationally interacting) body in any of the systems studied. The O-C diagrams indicate that the majority of the data are consistent with a linear ephemeris within the $\pm$3$ \sigma$ confidence level. We do however note the presence of a number of outliers beyond the $\pm$3$ \sigma$ limit for all of the systems studied. The use of data from different filters increases the likelihood of a larger dispersion in the mid-transit times because of the increased uncertainty in precisely defining the ingress and egress of the respective lightcurves.\\
\end{itemize}
\section*{Acknowledgements}
\noindent
This research is supported by the Adiyaman University Scientific Research Project Unit through the project number FEFMAP/2017-0002. We thank TUBITAK National Observatory for a partial support in using T100 telescope with project number 12CT100-388. The careful reading of our manuscript by the Referee helped to clarify a number of issues and is very much appreciated.  
\section*{Data availability}
The data underlying this article are available in the article and in its online supplementary material.
\bibliographystyle{mnras}

\begin{table*}
\caption{The Mid-Transit Times of HATP-36: 4-points from \citet{2012AJ....144...19B}; 26-points from \citet{2019AJ....157...82W},
 5-points from \citet{2015AandA...579A.136M}, 49-points from ETD, and 4-points from \citet{2019AJ....158...39C} }
\label{tab:midtimes_hatp36}
\begin{tabular}{lllll}
\hline\hline 
Cycle & BJD & $(O-C)$ (d) & error & References\\
\hline
-1391 & 2455555.891400 & 0.001800 & 0.0004870 &  \citet{2012AJ....144...19B}$^a$\\
-1360 & 2455597.037260 & -0.000108 & 0.0009980 &  \citet{2012AJ....144...19B}$^a$\\
-1357 & 2455601.020270 & 0.000860 & 0.0009550 &  \citet{2012AJ....144...19B}$^a$\\
-1351 & 2455608.984740 & 0.001246 & 0.0009860 &  \citet{2012AJ....144...19B}$^a$\\
-1071 & 2455980.639517 & -0.001229 & 0.0005500 &  ETD-Clear\\
-1065 & 2455988.603296 & -0.001534 & 0.0005000 &  ETD-R\\  
-1053 & 2456004.533406 & 0.000408 & 0.0005200 &  ETD-Clear\\
-1051 & 2456007.189090 & 0.001397 & 0.0007050 &  \citet{2019AJ....157...82W}\\
-1050 & 2456008.514756 & -0.000284 & 0.0002400 &  ETD-V\\  
-1045 & 2456015.151100 & -0.000677 & 0.0003430 &  \citet{2019AJ....157...82W}\\
-1035 & 2456028.425676 & 0.000426 & 0.0004900 &  ETD-Clear\\
-788 & 2456356.281780 & 0.001739 & 0.0010440 &  \citet{2019AJ....157...82W}\\
-783 & 2456362.916308 & -0.000470 & 0.0002400 &  ETD-I\\  
-781 & 2456365.568000 & -0.003473 & 0.0023400 &  \citet{2015AandA...579A.136M}\\
-776 & 2456372.208490 & 0.000281 & 0.0007850 &  \citet{2019AJ....157...82W}\\
-762 & 2456390.790978 & -0.000094 & 0.0005500 &  ETD-Clear\\
-757 & 2456397.421700 & -0.006109 & 0.0020200 &  \citet{2015AandA...579A.136M}\\
-757 & 2456397.427248 & -0.000561 & 0.0006500 &  ETD-Clear\\
-757 & 2456397.428940 & 0.001131 & 0.0003100 &  \citet{2015AandA...579A.136M}\\
-560 & 2456658.915591 & 0.000358 & 0.0004800 &  ETD-Clear\\
-543 & 2456681.479720 & -0.000417 & 0.0005800 &  ETD-R\\  
-519 & 2456713.335620 & -0.000853 & 0.0005900 &  ETD-Clear\\
-513 & 2456721.300850 & 0.000293 & 0.0008410 &  \citet{2019AJ....157...82W}\\
-510 & 2456725.281000 & -0.001599 & 0.0006760 &  \citet{2019AJ....157...82W}\\
-509 & 2456726.606939 & -0.003008 & 0.0006400 &  ETD-Clear\\
-507 & 2456729.264390 & -0.000251 & 0.0009510 &  \citet{2019AJ....157...82W}\\
-497 & 2456742.537889 & -0.000226 & 0.0006400 &  ETD-R\\  
-492 & 2456749.175800 & 0.000949 & 0.0010800 &  \citet{2019AJ....157...82W}\\
-489 & 2456753.158190 & 0.001297 & 0.0008460 &  \citet{2019AJ....157...82W}\\
-482 & 2456762.448340 & 0.000015 & 0.0001800 &  \citet{2015AandA...579A.136M}\\
-479 & 2456766.430550 & 0.000183 & 0.0002800 &  \citet{2015AandA...579A.136M}\\
-476 & 2456770.412089 & -0.000320 & 0.0006800 &  ETD-R\\  
-476 & 2456770.413029 & 0.000620 & 0.0006500 &  ETD-Clear\\
-465 & 2456785.014250 & 0.001021 & 0.0012750 &  \citet{2019AJ....157...82W}\\
-292 & 2457014.645585 & 0.001267 & 0.0005400 &  ETD-Clear\\
-259 & 2457058.446140 & -0.000640 & 0.0014890 &  \citet{2019AJ....157...82W}\\
-253 & 2457066.412020 & 0.001156 & 0.0007760 &  \citet{2019AJ....157...82W}\\
-250 & 2457070.392800 & -0.000106 & 0.0008580 &  \citet{2019AJ....157...82W}\\
-247 & 2457074.375934 & 0.000986 & 0.0008100 &  ETD-Clear\\
-201 & 2457135.432534 & -0.000391 & 0.0007300 &  ETD-R\\  
-199 & 2457138.086450 & -0.001169 & 0.0007330 &  \citet{2019AJ....157...82W}\\
-3 & 2457398.248300 & 0.000604 & 0.0000880 &  \citet{2019AJ....157...82W}\\
0 & 2457402.229480 & -0.000258 & 0.0004120 &  \citet{2019AJ....157...82W}\\
3 & 2457406.211790 & 0.000010 & 0.0009220 &  \citet{2019AJ....157...82W}\\
24 & 2457434.085800 & -0.000274 & 0.0013190 &  \citet{2019AJ....157...82W}\\
25 & 2457435.413400 & -0.000022 & 0.0014930 &  \citet{2019AJ....157...82W}\\
28 & 2457439.394010 & -0.001454 & 0.0009780 &  \citet{2019AJ....157...82W}\\
34 & 2457447.358490 & -0.001058 & 0.0005280 &  \citet{2019AJ....157...82W}\\
35 & 2457448.684605 & -0.002290 & 0.0007000 &  ETD-Clear\\
43 & 2457459.304940 & -0.000734 & 0.0009830 &  \citet{2019AJ....157...82W}\\
44 & 2457460.632005 & -0.001016 & 0.0006700 &  ETD-Clear\\
46 & 2457463.286385 & -0.001331 & 0.0005200 &  ETD-Clear\\
54 & 2457473.906675 & 0.000181 & 0.0004500 &  ETD-Clear\\
59 & 2457480.542900 & -0.000331 & 0.0013000 &  This work\\
59 & 2457480.543200 & -0.000031 & 0.0011000 &  This work-TUG\\
59 & 2457480.543396 & 0.000165 & 0.0006400 &  ETD-Clear\\
67 & 2457491.163310 & 0.001300 & 0.0007150 &  \citet{2019AJ....157...82W}\\
80 & 2457508.416426 & -0.001099 & 0.0003200 &  ETD-Clear\\
83 & 2457512.400976 & 0.001409 & 0.0010700 &  ETD-Clear\\
85 & 2457515.054050 & -0.000212 & 0.0010210 &  \citet{2019AJ....157...82W}\\
89 & 2457520.366510 & 0.002859 & 0.0001900 &  This work\\
\hline
\end{tabular}
\end{table*}

\begin{table*}
\contcaption{The Mid-Transit Times of HATP-36}
 \label{tab:continued}
 \begin{tabular}{lllll}
  \hline
 Cycle & BJD & $(O-C)$ (d) & error & References\\
  \hline
122 & 2457564.166080 & -0.000033 & 0.0021960 &  \citet{2019AJ....157...82W}\\
282 & 2457776.540800 & -0.000886 & 0.0014000 &  This work\\
303 & 2457804.417600 & 0.001620 & 0.0022000 &  This work\\
306 & 2457808.395400 & -0.002622 & 0.0017000 &  This work\\
344 & 2457858.836783 & -0.000437 & 0.0004500 &  ETD-V\\
348 & 2457864.145440 & -0.001170 & 0.0016290 &  \citet{2019AJ....157...82W}\\
352 & 2457869.453700 & -0.002299 & 0.0008600 &  This work\\
361 & 2457881.403284 & 0.001159 & 0.0006300 &  ETD-Clear\\
385 & 2457913.261140 & 0.002679 & 0.0009400 &  This work\\
536 & 2458113.688899 & 0.000991 & 0.0007500 &  ETD-Clear\\
563 & 2458149.527500 & 0.001214 & 0.0032000 &  This work\\
569 & 2458157.488471 & -0.001899 & 0.0006100 &  ETD-V\\  
569 & 2458157.492100 & 0.001730 & 0.0014000 &  This work\\
575 & 2458165.455070 & 0.000616 & 0.0000070 &  \citet{2019AJ....158...39C}\\
578 & 2458169.435141 & -0.001355 & 0.0004500 &  ETD-R\\  
614 & 2458217.221600 & 0.000600 & 0.0000080 &  \citet{2019AJ....158...39C}\\
621 & 2458226.512303 & -0.000128 & 0.0007100 &  ETD-Clear\\
624 & 2458230.496853 & 0.002380 & 0.0003500 &  ETD-Clear\\
635 & 2458245.094640 & -0.000654 & 0.0000070 &  \citet{2019AJ....158...39C}\\
669 & 2458290.225690 & 0.000587 & 0.0000070 &  \citet{2019AJ....158...39C}\\
811 & 2458478.710952 & 0.002528 & 0.0006000 &  ETD-Clear\\
820 & 2458490.656972 & 0.002422 & 0.0004200 &  ETD-Clear\\
850 & 2458530.474453 & -0.000517 & 0.0004900 &  ETD-V\\  
850 & 2458530.475453 & 0.000483 & 0.0004300 &  ETD-Clear\\
850 & 2458530.477143 & 0.002173 & 0.0004400 &  ETD-V\\
856 & 2458538.437434 & -0.001620 & 0.0006900 &  ETD-V\\
878 & 2458567.640925 & 0.000230 & 0.0005700 &  ETD-V\\  
1125 & 2458895.495126 & -0.000360 & 0.0006100 &  ETD-Clear\\
1128 & 2458899.478046 & 0.000518 & 0.0006000 &  ETD-Clear\\
1131 & 2458903.459776 & 0.000206 & 0.0004800 &  ETD-Clear\\
1140 & 2458915.404086 & -0.001610 & 0.0005200 &  ETD-R\\  
1146 & 2458923.370667 & 0.000887 & 0.0005200 &  ETD-V\\  
1149 & 2458927.351257 & -0.000565 & 0.0007200 &  ETD-Clear\\
1149 & 2458927.351687 & -0.000135 & 0.0007600 &  ETD-Clear\\
1158 & 2458939.297080 & -0.000868 & 0.0007200 &  This work\\
1162 & 2458944.606127 & -0.001210 & 0.0004900 &  ETD-Clear\\
1174 & 2458960.533898 & -0.001607 & 0.0005600 &  ETD-Clear\\
1186 & 2458976.463018 & -0.000655 & 0.0006000 &  ETD-Clear\\
1189 & 2458980.444600 & -0.001115 & 0.0023000 &  This work\\
1204 & 2459000.358500 & 0.002575 & 0.0011000 &  This work\\
1207 & 2459004.339640 & 0.001673 & 0.0009500 &  This work\\
1210 & 2459008.319100 & -0.000909 & 0.0013000 &  This work\\
1213 & 2459012.299480 & -0.002571 & 0.0009900 &  This work\\
1216 & 2459016.283500 & -0.000593 & 0.0022000 &  This work\\
\hline
\multicolumn{5}{l}{$^a$ Mid-transit times calculated by \citet{2019AJ....157...82W} from observations of  \citet{2012AJ....144...19B}}\\
\end{tabular}
\end{table*}

\begin{table*}
\caption{The Mid-Transit Times of HATP-56: 20-points from ETD and 1-point from \citet{2015AJ....150...85H}}
\label{tab:midtimes_hatp56}
\begin{tabular}{lllll}
\hline\hline 
Cycle & BJD & $(O-C)$ (d) & error & References\\
\hline
-422 & 2456553.616450 & 0.001827 & 0.00042 &  \citet{2015AJ....150...85H}\\
-122 & 2457390.864544 & 0.000512 & 0.00114 &  ETD-I\\
-115 & 2457410.400025 & 0.000173 & 0.00105 &  ETD-R\\
-110 & 2457424.351766 & -0.002243 & 0.00132 &  ETD-R \\
0 & 2457731.348700 & 0.003227 & 0.00130 &  This work\\
15 & 2457773.205700 & -0.002243 & 0.00200 &  This work\\
19 & 2457784.370992 & -0.000263 & 0.00109 &  ETD-R\\
24 & 2457798.328800 & 0.003375 & 0.00150&  This work\\
29 & 2457812.278900 & -0.000681 & 0.00160 &  This work\\
43 & 2457851.345304 & -0.005904 & 0.00103 &  ETD-I\\
114 & 2458049.501900 & 0.001657 & 0.00130 &  This work\\
133 & 2458102.528500 & 0.002462 & 0.00160 &  This work\\
138 & 2458116.481300 & 0.001105 & 0.00260 &  This work\\
143 & 2458130.432595 & -0.001749 & 0.00086 &  ETD-R\\
143 & 2458130.438100 & 0.003749 & 0.00300 &  This work\\
153 & 2458158.343800 & 0.001135 & 0.00280 &  This work\\
153 & 2458158.345516 & 0.002858 & 0.00081 &  ETD-V\\
156 & 2458166.712056 & -0.003096 & 0.00132 &  ETD-V\\
167 & 2458197.410617 & -0.003680 & 0.00128 &  ETD-V\\
287 & 2458532.312145 & -0.001916 & 0.00148 &  ETD-B\\
301 & 2458571.381665 & -0.004035 & 0.00135 &  ETD-V\\
391 & 2458822.555807 & -0.004716 & 0.00077 &  ETD-V\\
401 & 2458850.468847 & 0.000011 & 0.00134 &  ETD-Clear\\
411 & 2458878.372457 & -0.004693 & 0.00098 &  ETD-V\\
411 & 2458878.381967 & 0.004817 & 0.00053 &  ETD-Clear\\
416 & 2458892.330987 & -0.000320 & 0.0008 &  ETD-V\\
416 & 2458892.331847 & 0.000540 & 0.00104 &  ETD-Clear\\
416 & 2458892.333887 & 0.002580 & 0.00059 &  ETD-Clear\\
416 & 2458892.338547 & 0.007240 & 0.00125 &  ETD-Clear\\
435 & 2458945.357126 & 0.000023 & 0.00087 &  ETD-Clear\\
\hline
\end{tabular}
\end{table*}

\begin{table*}
\caption{The Mid-Transit Times of WASP-52: 89-points from ETD, 3-points from \citet{2013AandA...549A.134H}, 3-points from \citet{2015MNRAS.450.3101B}, 7-points from \citet{2019MNRAS.486.2290O}, 7-points from \citet{2017MNRAS.465..843M}, 1-point from \citet{2015JATIS...1b7002S}}
\label{tab:midtimes_wasp52}
\begin{tabular}{lllll}
\hline\hline 
Cycle & BJD & $(O-C)$ (d) & error & References\\
\hline
-1082 & 2455776.183950 & 0.000123 & 0.00015 &  \citet{2013AandA...549A.134H}$^a$\\
-1072 & 2455793.681180 & -0.000462 & 0.00022 &  \citet{2013AandA...549A.134H}$^a$\\
-1052 & 2455828.676800 & -0.000471 & 0.00015 &  \citet{2013AandA...549A.134H}$^a$\\
-799 & 2456271.373300 & 0.001321 & 0.00059 &  \citet{2015MNRAS.450.3101B}$^b$\\
-799 & 2456271.373781 & 0.001802 & 0.00042 &  ETD-Clear\\
-680 & 2456479.595130 & -0.000842 & 0.00016 &  \citet{2017MNRAS.465..843M}\\
-673 & 2456491.842899 & -0.001543 & 0.00019 &  ETD-Clear\\
-673 & 2456491.843240 & -0.001202 & 0.00026 &  \citet{2015MNRAS.450.3101B}$^b$\\
-660 & 2456514.593480 & 0.001879 & 0.00044 &  ETD-Clear\\
-645 & 2456540.839081 & 0.000758 & 0.00041 &  ETD-V\\
-644 & 2456542.589301 & 0.001197 & 0.00049 &  ETD-Clear\\
-640 & 2456549.588050 & 0.000820 & 0.00020 &  \citet{2017MNRAS.465..843M}\\
-635 & 2456558.335251 & -0.000887 & 0.00057 &  ETD-Clear\\
-635 & 2456558.336670 & 0.000532 & 0.00082 &  \citet{2015MNRAS.450.3101B}$^b$\\
-628 & 2456570.586342 & 0.001734 & 0.00023 &  ETD-V\\
-627 & 2456572.334632 & 0.000243 & 0.00068 &  ETD-Clear\\
-619 & 2456586.332182 & -0.000459 & 0.00049 &  ETD-Clear\\
-619 & 2456586.332930 & 0.000289 & 0.00012 &  \citet{2017MNRAS.465..843M}\\
-616 & 2456591.582273 & 0.000288 & 0.00047 &  ETD-Clear\\
-615 & 2456593.332583 & 0.000816 & 0.00044 &  ETD-Clear\\
-611 & 2456600.326963 & -0.003929 & 0.00077 &  ETD-Clear\\
-611 & 2456600.331783 & 0.000891 & 0.00045 &  ETD-Clear\\
-608 & 2456605.581703 & 0.001466 & 0.00050 &  ETD-Clear\\
-607 & 2456607.329813 & -0.000205 & 0.00049 &  ETD-R\\
-603 & 2456614.327303 & -0.001841 & 0.00077 &  ETD-Clear\\
-603 & 2456614.327803 & -0.001341 & 0.00071 &  ETD-R\\
-591 & 2456635.318764 & -0.007758 & 0.00076 &  ETD-V\\
-591 & 2456635.326324 & -0.000198 & 0.00042 &  ETD-Clear\\
-472 & 2456843.550270 & -0.000245 & 0.00060 &  ETD-V\\
-461 & 2456862.797710 & -0.000401 & 0.00006 &  \citet{2017MNRAS.465..843M}\\
-460 & 2456864.548610 & 0.000718 & 0.00063 &  ETD-V\\
-457 & 2456869.795640 & -0.001597 & 0.00053 &  ETD-Clear\\
-457 & 2456869.797690 & 0.000453 & 0.00034 &  ETD-Clear\\
-453 & 2456876.796050 & -0.000312 & 0.00011 &  \citet{2017MNRAS.465..843M}\\
-453 & 2456876.796640 & 0.000278 & 0.00057 &  ETD-Clear\\
-445 & 2456890.793930 & -0.000684 & 0.00018 &  \citet{2017MNRAS.465..843M}\\
-437 & 2456904.794361 & 0.001495 & 0.00058 &  ETD-Clear\\
-436 & 2456906.542370 & -0.000277 & 0.00008 &  \citet{2017MNRAS.465..843M}\\
-429 & 2456918.789620 & -0.001497 & 0.00039 &  \citet{2015JATIS...1b7002S}\\
-428 & 2456920.540622 & -0.000277 & 0.00069 &  ETD-R\\
-411 & 2456950.286413 & -0.000770 & 0.00042 &  ETD-Clear\\
-407 & 2456957.287113 & 0.000804 & 0.00031 &  ETD-V\\
-375 & 2457013.279514 & 0.000198 & 0.00025 &  ETD-V\\
-375 & 2457013.281124 & 0.001808 & 0.00052 &  ETD-R\\
-367 & 2457027.277924 & 0.000357 & 0.00038 &  ETD-V\\
-256 & 2457221.505008 & 0.001699 & 0.00030 &  ETD-Clear\\
-244 & 2457242.501708 & 0.001022 & 0.00020 &  ETD-Clear\\
-216 & 2457291.494279 & -0.000288 & 0.00040 &  ETD-R\\
-213 & 2457296.744149 & 0.000237 & 0.00052 &  ETD-Clear\\
-212 & 2457298.493359 & -0.000334 & 0.00057 &  ETD-Clear\\
-184 & 2457347.487359 & -0.000215 & 0.00062 &  ETD-Clear\\
-56 & 2457571.459900 & 0.000300 & 0.00060 &  \citet{2019MNRAS.486.2290O}\\
-52 & 2457578.458900 & 0.000174 & 0.00030 &  \citet{2019MNRAS.486.2290O}\\
-48 & 2457585.458900 & 0.001048 & 0.00040 &  \citet{2019MNRAS.486.2290O}\\
-40 & 2457599.456900 & 0.000797 & 0.00060 &  \citet{2019MNRAS.486.2290O}\\
-36 & 2457606.456077 & 0.000848 & 0.00040 &  ETD-Clear\\
-32 & 2457613.453397 & -0.000958 & 0.00042 &  ETD-Clear\\
-32 & 2457613.455430 & 0.001075 & 0.00064 &  This work\\
-24 & 2457627.452297 & -0.000309 & 0.00063 &  ETD-Clear\\
-24 & 2457627.452447 & -0.000159 & 0.00036 &  ETD-Clear\\
-24 & 2457627.452450 & -0.000156 & 0.00052 &  This work\\
-24 & 2457627.455800 & 0.003194 & 0.00180 &   \citet{2019MNRAS.486.2290O}\\
-24 & 2457627.458357 & 0.005751 & 0.00110 &  ETD-Clear\\
-21 & 2457632.702027 & 0.000076 & 0.00021 &  ETD-I\\
\hline
\end{tabular}
\end{table*}

\begin{table*}
\contcaption{The Mid-Transit Times of WASP-52}
 \label{tab:continued}
 \begin{tabular}{lllll}
 \hline
 Cycle & BJD & $(O-C)$ (d) & error & References\\
\hline
-20 & 2457634.451347 & -0.000385 & 0.00053 &  ETD-Clear\\
-20 & 2457634.451757 & 0.000025 & 0.00037 &  ETD-Clear\\
-20 & 2457634.452100 & 0.000368 & 0.00030 &  \citet{2019MNRAS.486.2290O}\\
-16 & 2457641.451177 & 0.000319 & 0.00073 &  ETD-Clear\\
-8 & 2457655.448567 & -0.000543 & 0.00062 &  ETD-Clear\\
-8 & 2457655.453917 & 0.004807 & 0.00059 &  ETD-Clear\\
-1 & 2457667.697047 & -0.000533 & 0.00029 &  ETD-Clear\\
-1 & 2457667.697617 & 0.000037 & 0.00037 &  ETD-R\\
0 & 2457669.446300 & -0.001061 & 0.00110 &  This work\\
0 & 2457669.450787 & 0.003426 & 0.00147 &  ETD-Clear\\
4 & 2457676.447417 & 0.000930 & 0.00046 &  ETD-Clear\\
8 & 2457683.445300 & -0.000313 & 0.00040 &  \citet{2019MNRAS.486.2290O}\\
31 & 2457723.690726 & 0.000140 & 0.00057 &  ETD-R\\
61 & 2457776.181600 & -0.002430 & 0.00170 &  This work\\
172 & 2457970.410691 & 0.000919 & 0.00162 &  ETD-Clear\\
188 & 2457998.407460 & 0.001185 & 0.00030 &  ETD-I\\
196 & 2458012.402020 & -0.002507 & 0.00125 &  ETD-Clear\\
200 & 2458019.403700 & 0.000048 & 0.00260 &  This work\\
207 & 2458031.652089 & -0.000034 & 0.00041 &  ETD-Clear\\
212 & 2458040.398340 & -0.002690 & 0.00071 &  This work\\
215 & 2458045.649309 & -0.001065 & 0.00042 &  ETD-R\\
215 & 2458045.650359 & -0.000015 & 0.00021 &  ETD-Clear\\
216 & 2458047.398590 & -0.001566 & 0.00078 &  This work\\
220 & 2458054.394760 & -0.004521 & 0.00100 &  This work\\
220 & 2458054.399009 & -0.000272 & 0.00033 &  ETD-R\\
235 & 2458080.644628 & -0.001375 & 0.00045 &  ETD-Clear\\
375 & 2458325.615078 & -0.000329 & 0.00050 &  ETD-Clear\\
387 & 2458346.611867 & -0.000917 & 0.00029 &  ETD-Clear\\
387 & 2458346.612657 & -0.000127 & 0.00032 &  ETD-V\\
391 & 2458353.611627 & -0.000283 & 0.00031 &  ETD-V\\
400 & 2458369.358700 & -0.001243 & 0.00110 &  This work\\
400 & 2458369.359226 & -0.000717 & 0.00038 &  ETD-Clear\\
404 & 2458376.359130 & 0.000061 & 0.00081 &  This work\\
411 & 2458388.607785 & 0.000246 & 0.00060 &  ETD-V\\
412 & 2458390.357705 & 0.000384 & 0.00044 &  ETD-Clear\\
412 & 2458390.361405 & 0.004084 & 0.00060 &  ETD-Clear\\
416 & 2458397.354120 & -0.002327 & 0.00088 &  This work\\
416 & 2458397.357025 & 0.000578 & 0.00060 &  ETD-Clear\\
424 & 2458411.353775 & -0.000923 & 0.00024 &  ETD-Clear\\
424 & 2458411.354215 & -0.000483 & 0.00064 &  ETD-Clear\\
432 & 2458425.353644 & 0.000694 & 0.00039 &  ETD-R\\
447 & 2458451.598023 & -0.001649 & 0.00049 &  ETD-Clear\\
571 & 2458668.572803 & 0.000231 & 0.00036 &  ETD-Clear\\
583 & 2458689.570531 & 0.000581 & 0.00041 &  ETD-Clear\\
591 & 2458703.568431 & 0.000230 & 0.00038 &  ETD-Clear\\
615 & 2458745.562929 & -0.000027 & 0.00032 &  ETD-Clear\\
623 & 2458759.565659 & 0.004451 & 0.00063 &  ETD-Clear\\
644 & 2458796.306037 & -0.000581 & 0.00043 &  ETD-V\\
656 & 2458817.303676 & -0.000320 & 0.00066 &  ETD-R\\
656 & 2458817.303776 & -0.000220 & 0.00025 &  ETD-Clear\\
660 & 2458824.303586 & 0.000464 & 0.00052 &  ETD-Clear\\
771 & 2459018.528617 & -0.000246 & 0.00076 &  ETD-Clear\\
795 & 2459060.524995 & 0.001377 & 0.00053 &  ETD-R\\
803 & 2459074.521354 & -0.000516 & 0.00053 &  ETD-Clear\\
807 & 2459081.520214 & -0.000781 & 0.00036 &  ETD-Clear\\
807 & 2459081.521044 & 0.000049 & 0.00035 &  ETD-R\\
819 & 2459102.517714 & -0.000659 & 0.00044 &  ETD-Clear\\
824 & 2459111.263700 & -0.003580 & 0.00120 &  This work\\
832 & 2459125.267100 & 0.001568 & 0.00110 &  This work\\
\hline
\multicolumn{5}{l}{$^a$ Mid-transit times calculated by \citet{2017MNRAS.465..843M} from observations of \citet{2013AandA...549A.134H}}\\
\multicolumn{5}{l}{$^b$ Mid-transit times calculated by \citet{2019MNRAS.486.2290O} from observations of \citet{2015MNRAS.450.3101B}}\\
\end{tabular}
\end{table*}
\appendix
\section{}
\begin{figure*}
	\includegraphics[width=\textwidth]{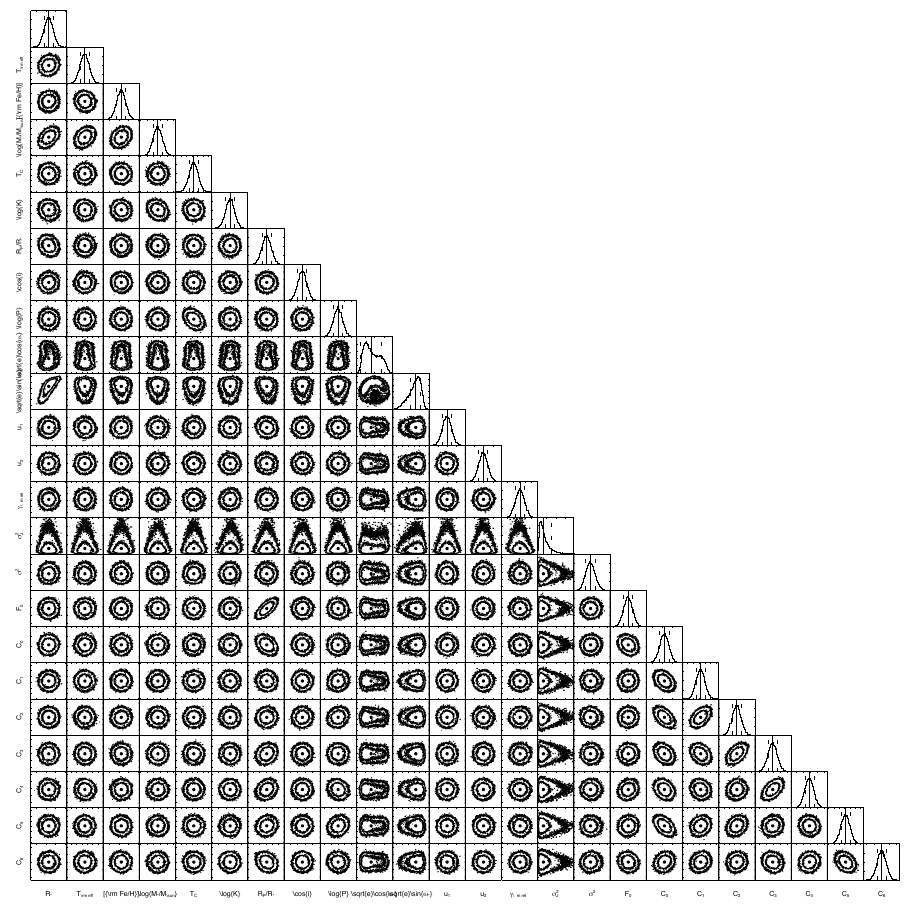}
    \caption{The posterior distributions obtained for the main stellar and planetary parameters of HATP-36 system.} 
    \label{fig:HATP-36b_covar}
\end{figure*}
\begin{figure*}
	\includegraphics[width=\textwidth]{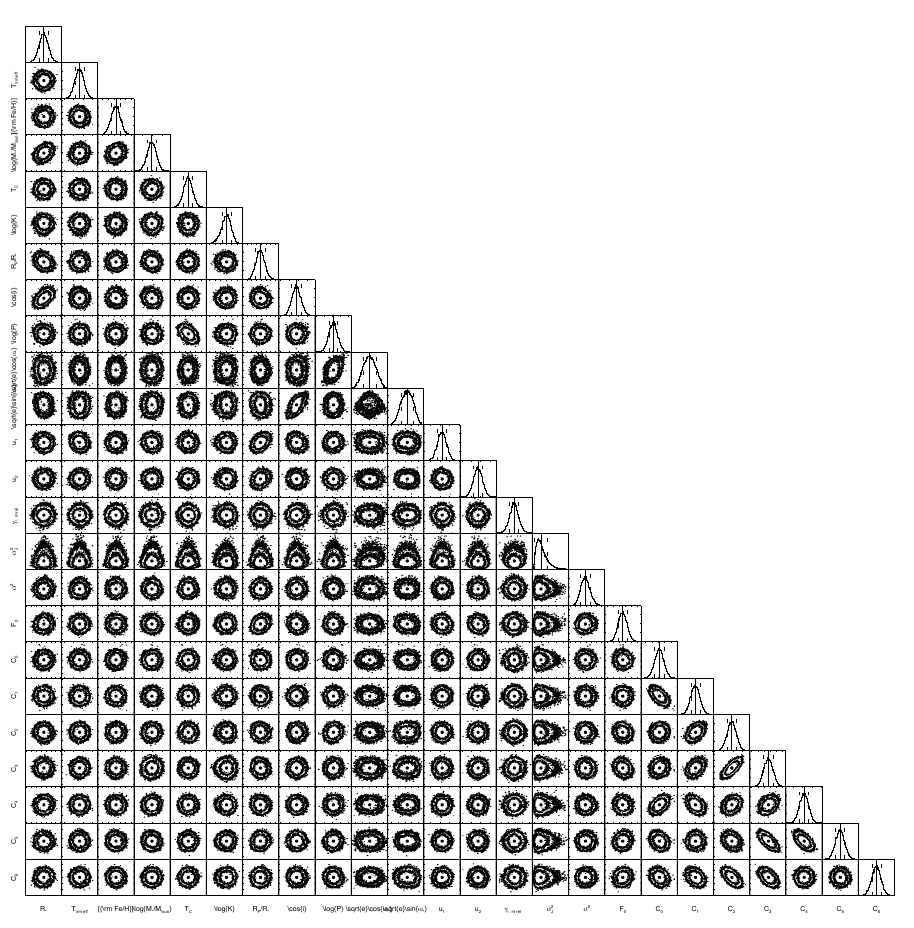}
    \caption{The posterior distributions obtained for the main stellar and planetary parameters of HATP-56 system.} 
    \label{fig:HATP-56b_covar}
\end{figure*}
\begin{figure*}
	\includegraphics[width=\textwidth]{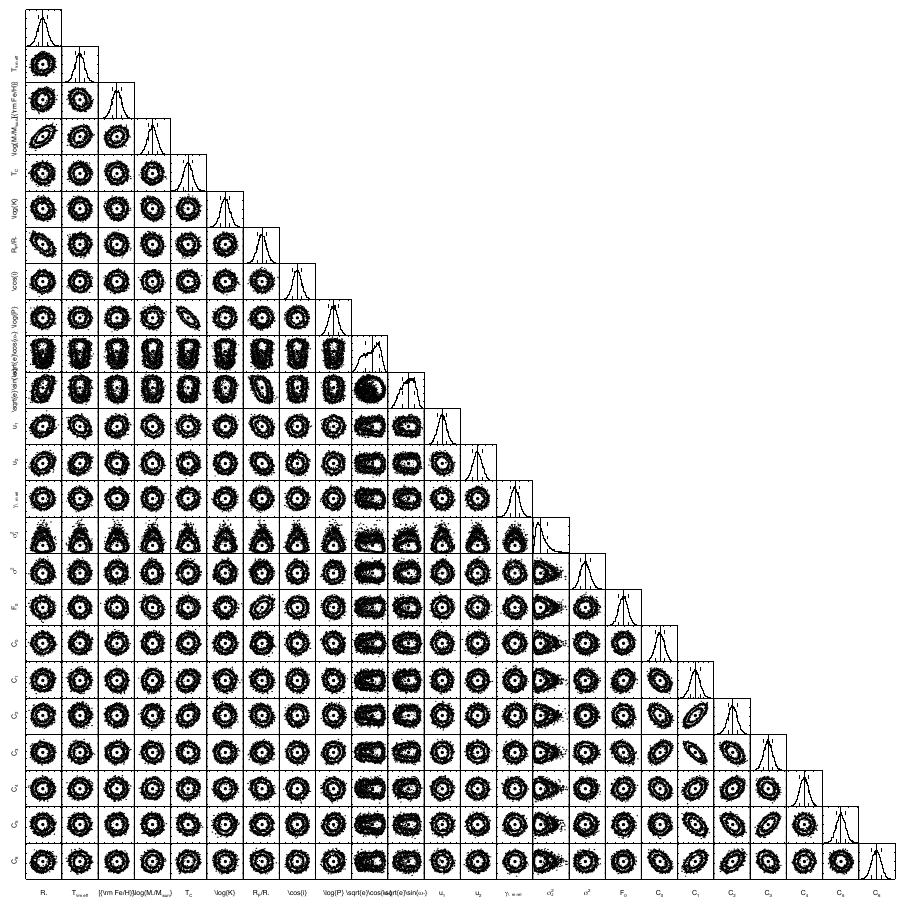}
    \caption{The posterior distributions obtained for the main stellar and planetary parameters of WASP-52 system.} 
    \label{fig:WASP-52b_covar}
\end{figure*}
\end{document}